\documentclass[useAMS,usenatbib]{mn2e}
\usepackage[english]{babel}
\usepackage{graphicx}
\usepackage{hyperref}
\usepackage{color}
\usepackage{amssymb,amsmath}
\usepackage{hhline}
\title[Photometric redshifts for the PAU Survey]{Precise photometric redshifts with a narrow-band filter set: The PAU Survey at the William Herschel Telescope}
\author[P. Mart\'{\i}, R. Miquel, F. J. Castander, E. Gazta\~{n}aga, M. Eriksen, C. S\'{a}nchez]{P. Mart\'{\i}$^{1}$\thanks{\href{mailto:pmarti@ifae.es}{E-mail: pmarti@ifae.es}}, R. Miquel$^{1,2}$, F. J. Castander$^{3}$, E. Gazta\~{n}aga$^{3}$, M. Eriksen$^{3}$
\newauthor
and C. S\'{a}nchez$^{1}$ \\
$^{1}$Institut de F\'{i}sica d'Altes Energies, Universitat Aut\`{o}noma de Barcelona, E-08193 Bellaterra (Barcelona), Spain\\
$^{2}$Instituci\'o Catalana de Recerca i Estudis Avan\c{c}ats, E-08010 Barcelona, Spain\\
$^{3}$Institut de Ci\`encies de l'Espai (ICE, IEEC/CSIC), E-08193 Bellaterra(Barcelona), Spain}

\begin{document}

\renewcommand{\refname}{REFERENCES}
\newcommand{\doctype}{paper}

\pagerange{\pageref{firstpage}--\pageref{lastpage}} \pubyear{2014}

\maketitle

\label{firstpage}

\begin{abstract}
The Physics of the Accelerating Universe (PAU) survey at the William
Herschel Telescope (WHT) will use a new optical camera (PAUCam) with a large set of
narrow-band filters to perform a photometric galaxy survey with a 
quasi-spectroscopic redshift precision of $\sigma(z)/(1+z) \sim 0.0035$ and map the
large-scale structure of the universe in three dimensions up to
$i_{AB} < 22.5$-$23.0$. 
In this paper we present a detailed photo-$z$ performance
study using photometric simulations for 40 equally-spaced 12.5-nm-wide
(FWHM) filters with a $\sim$25\% overlap and
spanning the wavelength range from 450~nm to 850~nm, together with a $ugrizY$ broad-band filter system.
We then present the migration
matrix $r_{ij}$, containing the probability
that a galaxy in a true redshift bin $j$ is measured in a photo-$z$ bin $i$, 
and study its effect on the determination of galaxy auto- and cross-correlations.
Finally, we also study the impact on the photo-$z$ performance of small variations of the filter set in terms of width, wavelength coverage, etc.,
and find a broad region where slightly modified filter sets provide
similar results, with
the original set being close to optimal.

\end{abstract}


\section{Introduction}

Galaxy surveys are a fundamental tool in order to understand the large-scale
structure of the universe as well as its geometry, content, history,
evolution and destiny.
Spectroscopic surveys~(2dF, \citet{Colless2001}; VVDS, \citet{LeFevre2005}; WiggleZ, \citet{Drinkwater2010}; BOSS, \citet{Dawson2013}) provide a 3D image of the
galaxy distribution in the near universe, but most of them 
suffer from limited depth, incompleteness and selection effects. 
Imaging surveys~(SDSS, \citet{York2000}; PanSTARRS, \citet{kaiser2000}; LSST, \citet{Tyson2003}) solve these problems but, on the other hand,
do not provide a true 3D picture of the universe, due to their limited
resolution in the position along the line of sight, 
which is obtained measuring the galaxy redshift through photometric
techniques using a set of broad-band filters. The Physics of the
Accelerated Universe (PAU) survey at the William Herschel Telescope
(WHT) in the Roque de los Muchachos Observatory (ORM) in the Canary island
of La Palma (Spain) will use narrow-band filters to try to achieve a
quasi-spectroscopic precision in the redshift determination that will
allow it to map the large-scale structure of the universe in 3D using
photometric techniques, and, hence, overcoming the limitations of
spectroscopic surveys~\citep{Benitez2009}.

In this paper we present the study of the photo-$z$ performance and its impact on clustering measurements expected
in the PAU survey in a sample consisting of all galaxies of all types
with $i_{AB} < 22.5$. Based of detailed simulation
studies~\citep{Gaztanaga2012}, the requirement for the precision is set at
$\sigma(z)/(1+z) = 0.0035$. The PAU survey will observe many galaxies beyond the
$i_{AB}=22.5$ limit that play a crucial role in
the PAU science case~\citep{Gaztanaga2012}. We will also study the performance 
in a fainter galaxy sample with $22.5 < i_{AB} \lesssim 23.7$,
expecting to reach a photo-$z$ precision not worse than $\sigma(z)/(1+z) = 0.05$~\citep{Gaztanaga2012}. Finally,
we will also study the impact on the photo-$z$ performance of small variations on the default filter set, in terms of width, wavelength coverage, etc.

There are two main sets of techniques for measuring photometric redshifts (or photo-$z$s): template methods~(e.g. \texttt{Hyperz}, \citet{Bolzonella2000}; \texttt{BPZ}, \citet{Benitez2000} \& \citet{Coe2006}; \texttt{LePhare}, \citet{Ilbert2006}; \texttt{EAZY}, \citet{Brammer2008}), in which the measured broadband galaxy spectral energy distribution (SED) is compared to a set of redshifted templates until a best match is found, thereby determining both the galaxy type and its redshift;
training methods~(e.g. \texttt{ANNz}, \citet{Collister2004}; \texttt{ArborZ}, \citet{Gerdes2010}; \texttt{TPZ}, \citet{CarrascoKind2013}), in which a set of galaxies for which the redshift is already known is used to train a machine-learning algorithm (an artifitial neural network, for example), which is then applied over the galaxy set of interest.
Each technique has its own advantages and disadvantages, whose discussion lies beyond the scope of this \doctype.

Throughout the paper we will be using the Bayesian Photo-Z ({\tt BPZ}) template-based
code from~\citet{Benitez2000}, after adapting it to our needs. We have also
tried several photo-$z$ codes based on training methods. We have found that, because of the
large, ${\cal O}(50)$, number of filters, some of them
run into difficulties due to the combinatorial growth of the
complexity of the problem, while others confirm the results
presented here. The results obtained with training methods will be described in
detail elsewhere (Bonnett et al., in preparation).

The outline of the paper is as follows.
In section~\ref{sec:filt} we present the default PAU filter set.
Section~\ref{sec:mock} discusses the mock galaxy samples that we use
in our study, the noise generation, and the split into a bright and a
faint galaxy samples. In section~\ref{sec:photoz} we introduce the 
{\tt BPZ} original code and our modifications, with special emphasis
on the prior redshift probability distributions and the
{\em odds} parameter. We also show the results obtained when running
{\tt BPZ} on the mock catalog using the default filter
set. Furthermore, we compute the so-called migration matrix $r_{ij}$
\citep{Gaztanaga2012}, corresponding to the probability that a galaxy at a true redshift bin
$j$ is actually measured at a photo-$z$ bin $i$, and its effect on the
measurement of galaxy auto- and cross-correlations.
In section~\ref{sec:opti} we try several modifications to the
filter set (wider/narrower filters, bluer/redder filters, etc.), study their performance on the brighter and fainter galaxy
samples and find the optimal set. 
Finally, in section~\ref{sec:discussion}, we discuss the
results and offer some conclusions.

\section{Default filter set-up}
\label{sec:filt}

In this section we construct the effective filter response $R(\lambda)$ of the PAU bands and compute their 5-$\sigma$ limiting magnitudes, $m_{AB}(5\sigma)$.

\subsection{Nominal response}
\begin{figure}
\centering
\includegraphics[width=85mm]{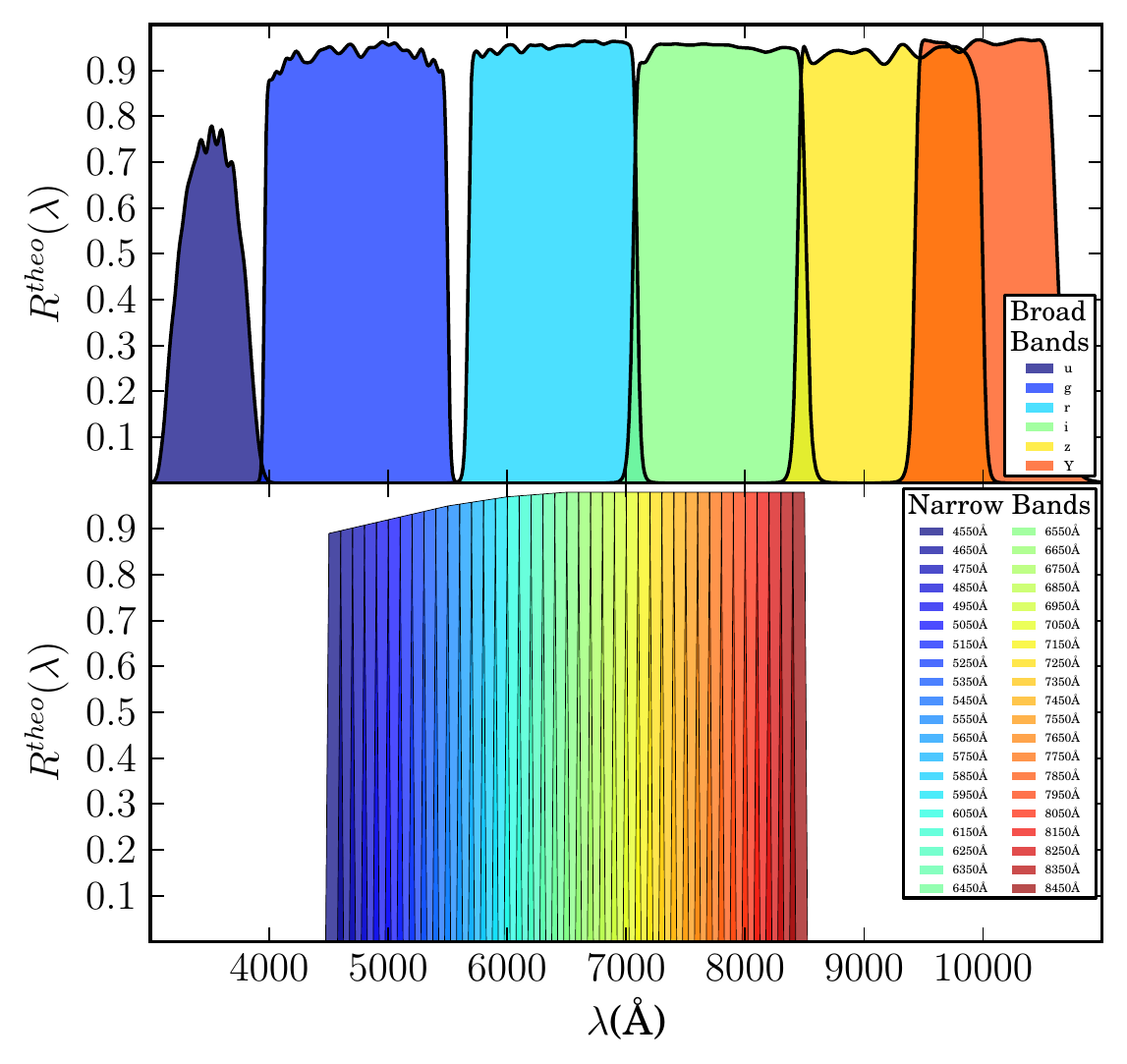}
\caption{The nominal (or theoretical) response $R^{theo}(\lambda)$ of the \textit{ugrizY} PAU Broad Bands (top) and the 40 Narrow Bands (bottom). The \textit{u} band is the same as in the USNO 40-in telescope at Flagstaff Station (Arizona), while the \textit{grizY} are the same as in the DECam mounted in the Blanco Telescope (CTIO, Chile). The narrow-band filters have a 125\AA\ {\em FWHM} and overlap by 25\% with the adjacent bands. They are labeled on the plot through their central wavelength. Their overall response is set to match that of the ALHAMBRA survey bands.}
\label{pau_theo_bands}
\end{figure}

PAUCam will mount two sets of filters: the Broad-Band~(BB) filters, composed of 6 bands \textit{ugrizY}\footnote{The $u$ band is assumed to be the same as the used in the USNO 40-in telescope at Flagstaff Station (Arizona) and its transmission can be obtained from \url{http://www.sdss.org/dr7/algorithms/standardstars/Filters/response.html}, while the rest are assumed to be the same as in the DECam mounted in the Blanco Telescope (CTIAO, Chile).}
, whose nominal (or theoretical) response $R^{theo}(\lambda)$ is shown on the top of Fig.~\ref{pau_theo_bands}, and the Narrow-Band~(NB) filters, shown on the bottom, which are composed of 40 top-hat adjacent bands with a rectangular width of 100\AA \ ranging from 4500\AA \ to 8500\AA. Since there is a technical limitation to construct such narrow top-hat bands, we relax the transition from 0 to the maximum response by adding two lateral wings of 25\AA \ width, resulting in a \textit{FWHM} of 125\AA. This induces an overlap of $\sim$25\% between contiguous bands. Additionally, we set the overall NB response to match that from the ALHAMBRA survey instrument \citep{Moles2008}, which are comparable in technical specifications (although with a wider wavelength range, $\sim 310$\AA, transmission) and have similar coatings.

\subsection{Effective response}
\begin{figure}
\centering
\includegraphics[width=80mm]{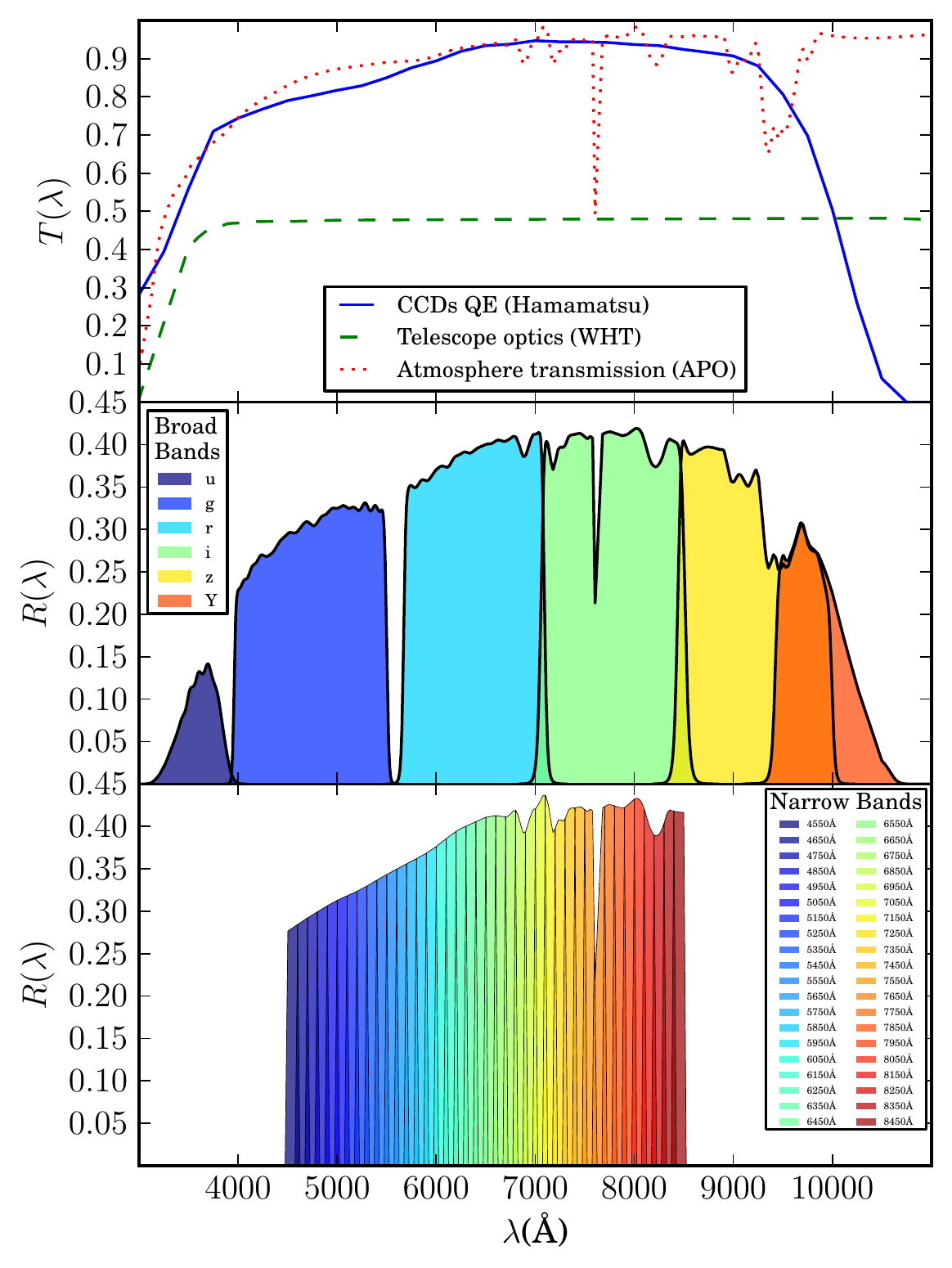}
\caption{Top: the QE curve (blue) of the PAUCam CCDs, the transmission curve of the WHT optics (green), and the atmospheric transmission (red) at the APO (Apache Point Observatory), that affect the final response of the PAU bands. The two lower plots are the same as in Fig.~\ref{pau_theo_bands}, but after taking into account these additional transmission curves $T_i(\lambda)$ through Eq.~(\ref{eff_filt}). For the sake of clarity, we have rescaled the $y$-axes due to the low efficiency of the mirror's reflexion.}
\label{pau_effective_bands}
\end{figure}

The filter responses $R^{theo}(\lambda)$ in Fig. \ref{pau_theo_bands} are the nominal: this is the response that we would measure if light went only through the filter. Light also goes through the Earth's atmosphere, which absorbs part of the light, and then, also goes through the optics (mirror and corrector) of the telescope before getting into the filter. Moreover, the CCD detectors behind filters also are affected by a Quantum Efficiency (QE) response curve. Therefore, if we want to know the effective response of the filters, we will have to take into account all the transmission curves $T_i(\lambda)$ of these effects $i$. In our case, these curves are shown in the top plot of Fig. \ref{pau_effective_bands}. The QE curve (blue) corresponds to the measured QE of CCDs provided by Hammamatsu, the measured transmission curve of the telescope's optics (mirror + corrector) (green) corresponds to that from the William Herschel Telescope (WHT) optics, and the atmospheric transmission curve (red) is taken from the Apache Point Observatory (APO) at New Mexico. We assume that the APO atmosphere transmission is close enough to the ORM for the purpose of this study. The resulting effective response $R(\lambda)$ is derived with the expression:
\begin{eqnarray}
R(\lambda) &=& R^{theo}(\lambda) \prod_{i} T_i(\lambda) \nonumber \\
&=& R^{theo}(\lambda) \cdot T_{CCD}(\lambda) \cdot T_{opt}(\lambda) \cdot T_{atm}(\lambda).
\label{eff_filt}
\end{eqnarray}
The transmission of the WHT optics is less than 50\% in all the wavelength range, so that the resulting effective responses are significantly reduced. On the other hand, the three transmission curves $T_i(\lambda)$ begin to fall when they enter the ultraviolet region ($\sim$3800\AA). Similarly, the CCDs QE drops as we approach the infrared region above $\sim$9000\AA. Overall, the $u$ and $Y$ broad bands are less efficient than the rest. This does not affect the NB, since their wavelength range are within these limits. Atmospheric telluric absorption bands, located between $\sim$700nm and $\sim$1$\mu$m, are also imprinted in the final response of the filters. This is particularly relevant for the NB since their typical width is similar to the width of these valleys. In particular, the profile of the narrow band with central wavelength at $\sim$7550\AA \ (orange) is drastically changed by the telluric absorption $A$-band. 

\subsection{5-$\sigma$ limiting magnitudes}
\begin{figure}
\centering
\includegraphics[width=80mm]{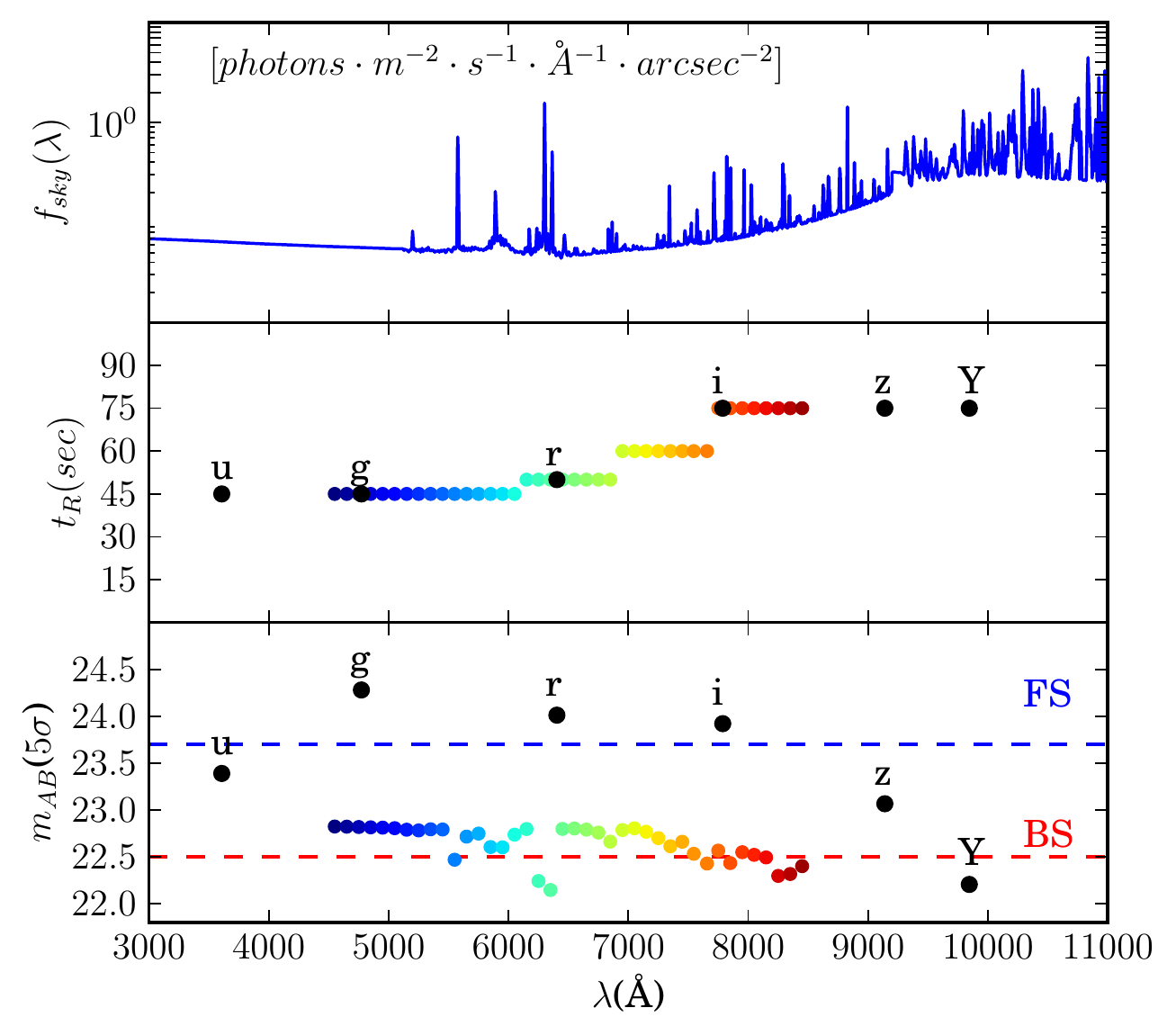}
\caption{Top: a model of the spectral density flux of the sky brightness $f_{sky}(\lambda)$ at La Palma (day 7 on the lunar cycle) assuming an airmass of 1.0, used in (\ref{Nsky}). Middle: the exposure times $t_R$ for each PAU band used in (\ref{Ngal}) and (\ref{Nsky}). Bottom: the resulting limiting magnitudes $m_{AB}(5\sigma)$ for each band computed through (\ref{SN}), (\ref{Ngal}) and (\ref{Nsky}). Colored points correspond to Narrow Bands and black to Broad Bands.}
\label{pau_lim_mag}
\end{figure}

Next, we compute the 5-$\sigma$ limiting magnitudes, $m_{AB}(5\sigma)$, for all the PAU bands in the AB photometric system\footnote{According to \citet{Hogg1996}, the apparent magnitude $m_{AB}$ in the AB system in a band with response $R(\lambda)$ for a source with spectral density flux $f(\nu)$ (energy per unit time per unit area per unit frequency) is defined as $m_{AB} \equiv-2.5\log_{10}\left[\int f(\nu) R(\nu){d\nu \over\nu} / \int \text{(3631Jy)} R(\nu){d\nu \over\nu}\right]$, where $1Jy = 10^{-23} erg\cdot s^{-1} \cdot cm^{-2} \cdot Hz^{-1}$ or $1.51 \cdot 10^7 photons\cdot m^{-2} \cdot s^{-1} \cdot {\lambda \over d\lambda}$ in wavelength space.} \citep{Oke1970}. This is the apparent magnitude whose Signal-to-Noise ratio, given by
\begin{equation}
{S \over N} = \sqrt{A \over \alpha^2}{N_{gal} \over \sqrt{N_{gal} + N_{sky} + n RN^2} },
\label{SN}
\end{equation}
is equal to 5, where 
\begin{eqnarray}
N_{gal} &=& 3631 \cdot 1.51\cdot10^7  \cdot 10^{-0.4m_{AB}} \cdot \left({\alpha^2\over A}\right) \cdot \nonumber \\
&\cdot& \pi\left({\phi \over 2}\right)^2 \cdot nt_R \cdot \int^{\infty}_0 R(\lambda) {d\lambda \over \lambda} \label{Ngal}, \label{Ngal} \\
N_{sky} &=& \alpha^2 \cdot \pi\left({\phi \over 2}\right)^2 \cdot nt_R \cdot \int^{\infty}_0 f_{sky}(\lambda) R(\lambda) d\lambda, \label{Nsky}
\end{eqnarray}
are the photons per pixel coming from the galaxy and the sky brightness  respectively, $\lbrace \phi, \alpha, RN, A, n \rbrace$ are the parameters of both the WHT and PAUCam instrument, whose values and description are given in Table \ref{PAU_parameters}, $f_{sky}(\lambda)$ is the spectral density flux per unit of aperture of the sky brightness, whose curve is on top of Fig.~\ref{pau_lim_mag}, and $t_R$ is the exposure time for the filter $R(\lambda)$. 
\begin{table}
\caption{Description and values of the WHT and PAUCam parameters used in (\ref{SN}), (\ref{Ngal}) and (\ref{Nsky}), to compute the Signal-to-Noise ratio ($S/N$).}
\vspace*{12pt}
\centering
\begin{tabular}{|c|c|c|}
\hline
 $\phi$ & Telescope mirror diameter & 4.2 m \\
$\alpha$ & Focal Plane Scale & 0.265 arcsec/pix \\
$RN$ & Read-out Noise & 5 electrons/pix \\
$A$ & Galaxy Aperture & 2 arcsec$^2$ \\
$n$ & \# of Exposures & 2 \\
\hline
\end{tabular}
\label{PAU_parameters}
\end{table}
All the filters intended for the photometry are arranged over the central part of the Focal Plane (FP) where vignetting is practically negligible. NB are distributed through 5 interchangeable trays. From the bluest to the reddest, each tray carries a group of 8 consecutive NB. This gives 5 trays $\times$ 8 NB $=$ 40 NB. 
Values for the exposure times $T_i$ of each tray are shown on the left column of Table \ref{PAU_Texp}. On the other hand, each BB filter is mounted into a dedicated tray with its particular exposure time, in such a way that NB and BB exposure times are completely decoupled. Values for the exposure times $t_R$ of each BB filter are shown on the right column of Table \ref{PAU_Texp}.
\begin{table}
\caption{Left: Exposure times $T_i$ for each PAUCam NB filter tray. The individual NB exposure times are equal to those of the tray where they are. Right: The BB exposure times. Exposure times $t_R$ per filter are also shown in the middle plot of Fig.~\ref{pau_lim_mag}.}
\vspace*{12pt}
\centering
\begin{tabular}{|c|c|c|c|c|}
\multicolumn{2}{c}{NB tray $T_i$} & & \multicolumn{2}{c}{BB $t_R$} \\
\cline{1-2} \cline{4-5} 
$T_1$  & 45 sec & & u & 45 sec \\
$T_2$  & 45 sec & & g & 45 sec \\
$T_3$  & 50 sec & & r & 50 sec \\
$T_4$  & 60 sec & & i & 75 sec \\
$T_5$  & 75 sec & & z & 75 sec \\
       &        & & Y & 75 sec \\
\cline{1-2} \cline{4-5} 
\end{tabular}
\label{PAU_Texp}
\end{table}
The exposure times $t_R$ and the derived limiting magnitudes $m_{AB}(5\sigma)$ for each filter are also shown on the middle and bottom plots of Fig.~\ref{pau_lim_mag} respectively, in a color degradation for NB and in black for BB. 
Since $f_{sky}(\lambda)$ increases with wavelength, exposure times $t_R$ for redder filters are also set to increase in order to compensate the noise introduced by the sky (note that the exposure times for the NB increase in steps, due to their arrangement in groups per tray). However, this increment is not enough to compensate the sky brightness as we can see with the descending $m_{AB}(5\sigma)$ with wavelength. Note that the $u$ band has a lower limiting magnitude compared with $g$ even being at shorter wavelengths. This is because the $u$ response is strongly affected by the transmission curves $T_i$. Furthermore, there are large drops in $m_{AB}(5\sigma)$ for the NB with central wavelength 5550\AA, 6250\AA \ and 6350\AA. This is  due to emission lines in the sky spectrum $f_{sky}(\lambda)$ at these wavelengths.

\section{The mock catalog}
\label{sec:mock}

In this section we generate a photometric mock catalog $\lbrace m_j \pm \sigma_{m_j}, z,t \rbrace$ with observed magnitudes $m_j\pm\sigma_{m_j}$ in each PAU band $j$ for galaxies at redshift~$z$ and with spectral type~$t$.

\subsection{Noiseless magnitudes}
We use a method similar to the one described in \citet{Jouvel2009}, which consists on sampling the cumulative Luminosity Function (LF): 
\begin{eqnarray}
N(z,t) = \int_{-\infty}^{M_{lim}(z,t)} \phi x^\alpha e^{-x} dx \nonumber \\
x\equiv10^{-0.4(M-M^*)},
\label{z_dist}
\end{eqnarray}
in the redshift range $z=[0,6]$, for a total of $N\sim10^6$ galaxies. $M$~is the absolute magnitude and $M_{lim}(z,t)$ the absolute magnitude limit at redshift $z$ and spectral type $t$ for a given apparent magnitude limit of the catalog $m_{lim}$ in some reference band. In our case $m_{lim}<26.0$ in the $r_{SDSS}$ band. Finally, $\lbrace M^*, \phi, \alpha \rbrace$ are the parameters of the LF, which also depend on $z$ and $t$. We assume that their redshift dependency is:
\begin{equation}
 \lbrace M^*, \log_{10}\phi, \alpha \rbrace = \lbrace a \exp[-(1+z)^{b}] + c \rbrace
\label{LF_param_evolv}
\end{equation}
where $\lbrace a, b, c \rbrace$ are type-dependent parameters whose values are in Table~\ref{LF_parameters}. These values are chosen to match the LFs and their evolution from \cite{Dahlen2005}, where three different spectral types, 1 = Elliptical (Ell), 2 = Spiral (Sp), 3 = Irregular (Irr), are distinguished. 
\begin{table*}
\centering
\caption{Parameters $\lbrace a,b,c \rbrace$ that, through (\ref{LF_param_evolv}), give the values and evolution in redshift of the LF parameters $\lbrace \log_{10}\phi, M^*, \alpha \rbrace$ for a given spectral type $t$. These values are based on the LFs in \citet{Dahlen2005}, where three galaxy types are distinguished: 1 = Ell, 2 = Sp and 3 = Irr.}
\vspace*{12pt}
\begin{tabular}{c|ccc|ccc|ccc|}
\cline{2-10}
& \multicolumn{3}{c|}{$\log_{10}\phi$} & \multicolumn{3}{c|}{$M^*$} & \multicolumn{3}{c|}{$\alpha$}\\
\hline
\multicolumn{1}{|c|}{t} & a & b & c & a & b & c & a & b & c\\
\hline
\multicolumn{1}{|c|}{1} & 2.4 & 1.1 & -2.7 & 5.0 & 1.6 &-21.90 & 1.7 & 1.6 & -1.00 \\
\multicolumn{1}{|c|}{2} & 0.5 & 0.1 & -2.28 & 3.2 & 2.5 & -21.00 & 0.7 & -0.9 & -1.50 \\
\multicolumn{1}{|c|}{3} & 1.0 & -3.5 & -3.1 & 5.0 & 1.3 & -20.00 & 1.8 & 0.9 & -1.85 \\
\hline
\end{tabular}
\label{LF_parameters}
\end{table*}

\begin{figure}
\centering
\includegraphics[width=80mm]{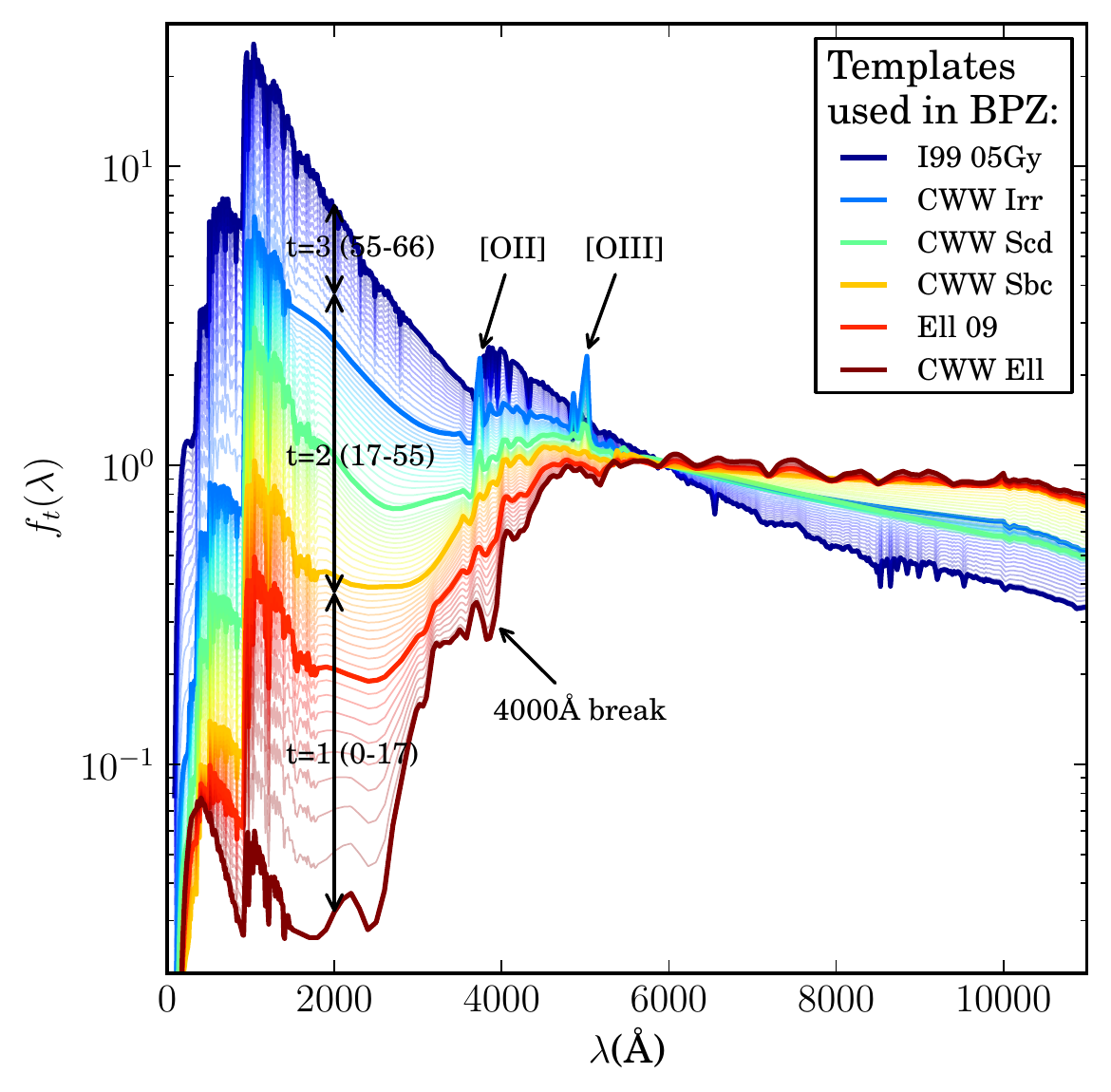}
\caption{The 66 spectral templates in the rest frame extracted from the extended CWW library. They are used, in (\ref{abs_m}) and (\ref{mi}), to generate the photometry of the PAU mock catalog. They evolve from Ellipticals (0-17, in red) to Spirals (17-55, in yellow, green and cyan), and finally to Irregulars (55-66, in blue and violet). Wider and deeper curves highlight the five templates used in \texttt{BPZ} to compute the photo-$z$s.}
\label{pau_templates}
\end{figure}

The relation between the absolute magnitude $M$ and the apparent magnitude $m$ for a galaxy at redshift $z$ and with spectral type $t$, used in the magnitude limit of (\ref{z_dist}), is:
\begin{equation}
M =  m - 5 \log_{10}D_L(z) - 25 - K(z,t)
\label{abs_m}
\end{equation}
where $D_L(z)$ is the luminosity distance of the galaxy at redshift $z$ in Mpc, which in a $\Lambda$CDM universe is expressed as:
\begin{equation}
D_L(z) \equiv (1+z) {c \over H_0}\int_0^{z} {dz \over \sqrt{\Omega_M(1+z)^3 + \Omega_\Lambda} },
\end{equation}
with cosmological parameters choosen to be: $H_0$ = 75~(km/s)/Mpc, $\Omega_M$ = 0.25 and $\Omega_\Lambda$ = 0.75, and $K(z,t)$ is the $K$-correction:
\begin{equation}
K(z,t) \equiv -2.5 \log_{10} \left[{1 \over 1+z}{\int^{\infty}_{0} f_t(\lambda) R_0(\lambda)\lambda d\lambda  \over \int^{\infty}_{0} f_t((1+z)\lambda)R_0(\lambda)d\lambda}\right]
\end{equation}
where $R_0(\lambda)$ is the response of the reference band, $f_t(\lambda)$ is the Spectral Energy Density (SED) of the galaxy with spectral type $t$ in the rest frame, and $f_t((1+z)\lambda)$ the same SED at redshift $z$. As a representation of these SEDs, we use the \texttt{CWW} \citep{Coleman1980} extended template library from the \texttt{LePhare}\footnote{The extended \texttt{CWW} library can be found in the folder \tt{/lephare\_dev/sed/GAL/CE\_NEW/} of the \texttt{LePhare} package at \url{http://www.cfht.hawaii.edu/~arnouts/LEPHARE/DOWNLOAD/lephare\_dev\_v2.2.tar.gz}} photo-$z$ code. It contains 66 templates ranging through Ell$\rightarrow$Sp$\rightarrow$Irr and shown in Fig.~\ref{pau_templates}. We split them in three groups: Ell = (0-17), Sp = (17-55) and Irr = (55-66). Then, a specific template within one of these groups is randomly selected and assigned to the galaxy. Actually, we allow the spectral type $t$ to range from 1 to 66 with a resolution of 0.01 by interpolating between templates.

After assigning $\lbrace z,t \rbrace$ values for all galaxies, we also assign them an absolute magnitude $M$ randomly within the range $[-\infty, M_{lim}(z,t)]$ following the LF probability distribution in (\ref{z_dist}). Then, the apparent magnitude $m_0$, in our reference band $r_{SDSS}$, is computed from (\ref{abs_m}). The other magnitudes $m_j$ at any band $j$ are obtained through:
\begin{equation}
m_j = m_0 + 2.5 \log_{10} \left[ {\int^{\infty}_{0} f_t((1+z)\lambda) R_0(\lambda)\lambda d\lambda \over \int^{\infty}_{0} f_t((1+z)\lambda) R_j(\lambda)\lambda d\lambda} {\int^{\infty}_{0} R_j(\lambda) {d\lambda \over \lambda} \over \int^{\infty}_{0} R_0(\lambda) {d\lambda \over \lambda}}\right]
\label{mi}
\end{equation}
where $R_j(\lambda)$ is the response of some PAU band $j$.

\subsection{Noisy magnitudes}
The resulting magnitudes are noiseless, so we have to transform them to observed magnitudes by adding a random component of noise as follows:
\begin{equation}
m_j \rightarrow m_j + \eta(0,1)\sigma_{m_j},
\label{noiselesstonoisy}
\end{equation}
where $\eta(0,1)$ is a normal variable and $\sigma_{m_j}$ the expected magnitude error which is related to the signal-to-noise in Eq.~\ref{SN} as follows:
\begin{equation}
\sigma_{m_j} = 2.5 \log_{10} \left(1 + {1 \over (S/N)_j} \right)
\label{errm}
\end{equation}
Additionally, we add an extra component of noise of size $\sim$0.022, corresponding to a $S/N=50$, in quadrature to $\sigma_m$, which takes into account some possible photometric calibration issues. Finally, we obtain the mock catalog $\lbrace m_j \pm \sigma_{m_j}, z,t \rbrace$. 

\begin{figure}
\centering
\includegraphics[width=70mm]{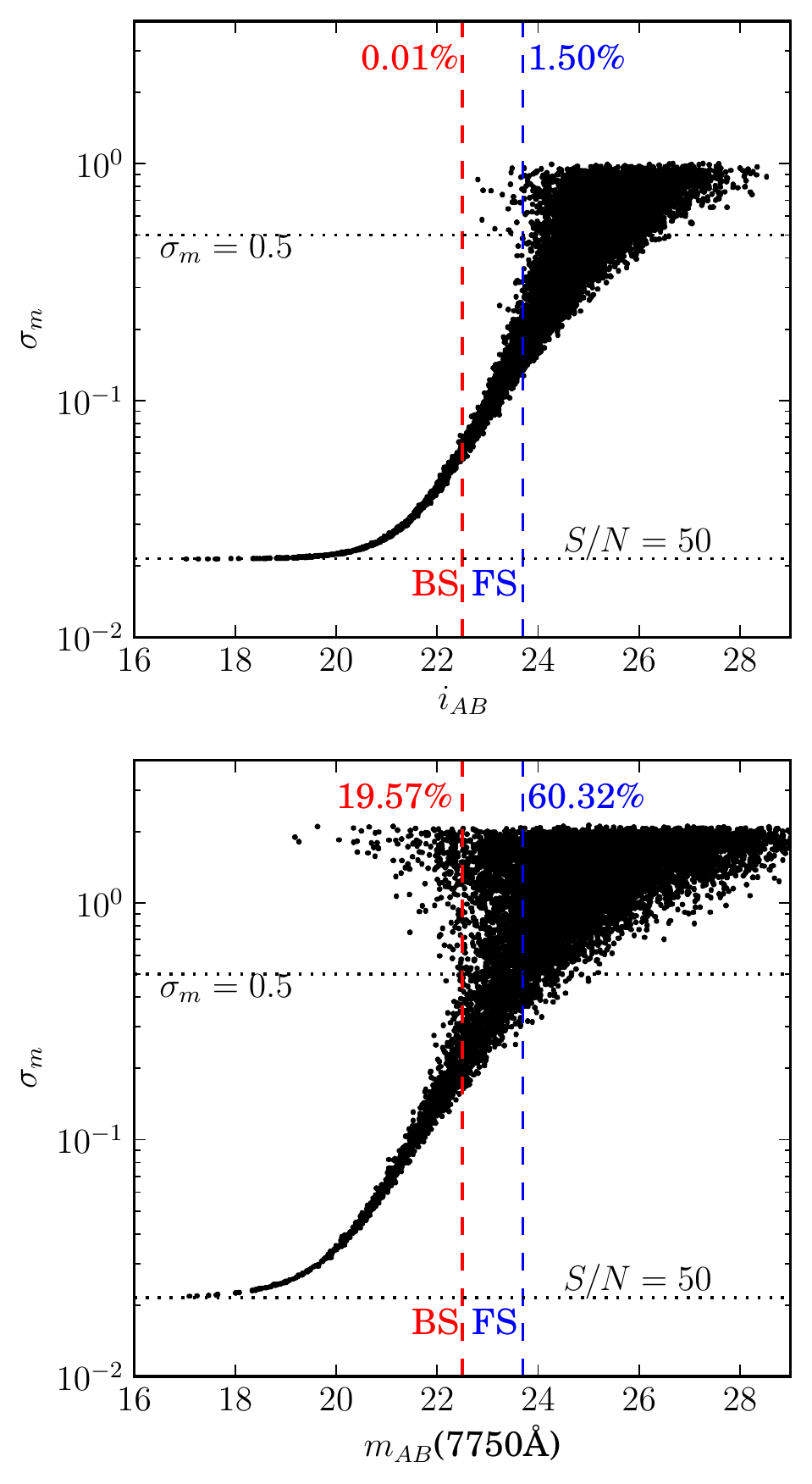}
\caption{Scatter plots of $\sigma_m$ vs. $m$ for the $i$ BB (top) and the 7750\AA \ NB (bottom). For the sake of clarity only 10000 randomly selected galaxies are plotted. The magnitude limits of the BS (red) and FS (blue) are also plotted as vertical-dashed lines. The bottom dotted line in both plots shows the calibration error ($S/N=50$) added in quadrature to $\sigma_m$, while the top dotted line shows the threshold where magnitudes are considered as non-observed. The proportions of non-observed magnitudes ($\sigma_m>0.5$) in each sample are also shown at the top with their correspondent color.}
\label{err_m}
\end{figure}

The resulting $\sigma_m$ vs. $m$ scatter plots, in the $i$ BB and the 7750\AA \ NB, are shown in Fig.~\ref{err_m}, where for the sake of clarity we only plot 10000 randomly selected galaxies. The 7750\AA \ band is chosen because its central wavelength is very similar to that of the $i$ band. Note how on both bands $\sigma_m$ starts being flat at $\sim$0.022 (the calibration error), and then, at fainter magnitudes, when the sky brightness and the CCD read-out noise become important, it grows and the scatter becomes wider. 

\subsection{Bright and Faint Samples}
The PAU survey science will be mostly focused on Large Scale Structure (LSS) studies such as cross measurements of Redshift Space Distortions (RSD) and Magnification Bias (MAG) between two galaxy samples: the Bright Sample (BS) on the foreground and the Faint Sample (FS) on the background (see \cite{Gaztanaga2012}). The BS should contain galaxies bright enough to have a large signal-to-noise in all bands, including the narrow bands, and, therefore, reach the necessary photo-$z$ accuracy to measure RSD. 
We see in Fig.~\ref{pau_lim_mag} that the 5-$\sigma$ limiting magnitudes for the NB are close to 22.5, so we define the BS as all those galaxies with $i_{AB} \equiv m_{AB}(i)<22.5$. The FS will contain the rest of the galaxies within $22.5<i_{AB}<23.7$. The upper limit has been chosen to roughly match the 5-$\sigma$ limiting magnitudes of the broad bands (see Fig.~\ref{pau_lim_mag}). 
We consider that a magnitude is not observed in one band if its correspondent error is $\sigma_{m}>0.5$. We find that a $\sim$0.01\% of galaxies in the BS are not observed in the $i$ band, with the fraction increasing to $\sim$1.5\% in the FS. Similarly, in the BS $\sim$19.57\% are not observed in the 7750\AA \ band, increasing to $\sim$60.32\% in the FS (see Fig.~\ref{err_m}). This tells us that, while most of the BB information will be present in both samples, the presence of NB information in the FS will be rather limited, degrading considerably the photo-$z$s. 

\begin{figure*}
\centering
\includegraphics[width=130mm]{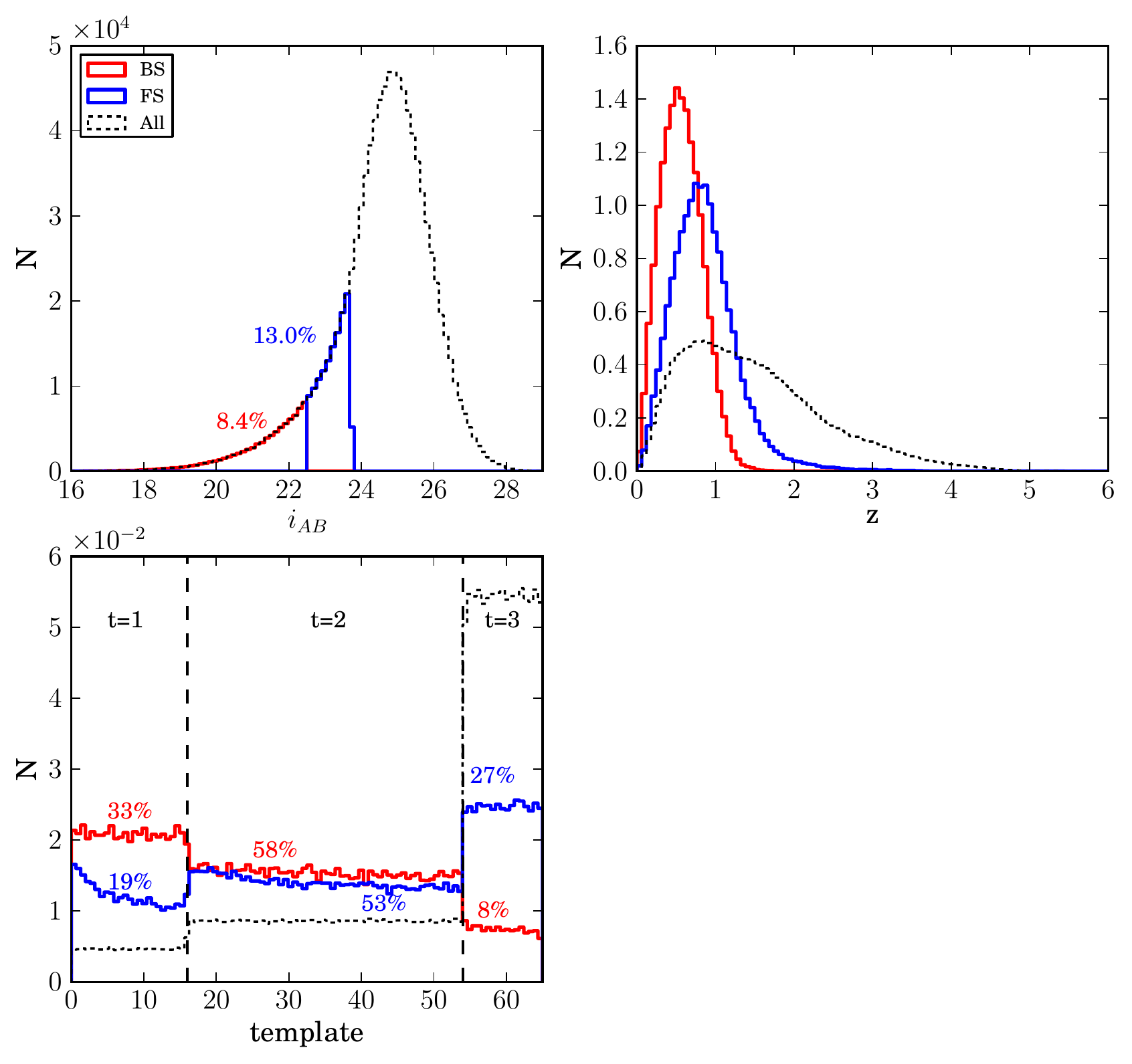}
\caption{The observed magnitude in the $i$ band (top-left), true redshift (top-right) and spectral type distributions for the whole catalog (black), the BS $i_{AB}<22.5$ (red) and the FS $22.5<i_{AB}<23.7$ (blue). We also show the fraction of the total galaxies in each sample in the magnitude distribution plot, in their corresponding color. Similarly, we show the proportions of spectral types in each sample. The redshift and type distributions have been normalized.}
\label{mzt_dist}
\end{figure*}

In Fig.~\ref{mzt_dist} we show the resulting distributions of the magnitude $i_{AB}$ (top-left), the true redshift $z$ (top-right) and the spectral type $t$ (bottom) of the galaxies in the whole catalog (black-dotted), the BS (red-solid) and the FS (blue-solid). The magnitude distribution of the whole catalog has its maximum at $\sim$25.0, so that the BS and FS are on the brighter tail of the distribution and account for $\sim$8.4\% and $\sim$13\% of the whole catalog respectively. However, this also helps both samples to have a very good completeness up to their magnitude limit. We can also see that, while the whole catalog extends up to $z\sim5$, the BS only goes up to $z\sim1.5$ and the FS up to $z\sim3$. Finally, we see that both BS and FS have a similar proportion of Spiral galaxies ($t=2$), $\sim55$\%; however the BS contains more elliptical galaxies (33\%) than the FS (19\%), and consequently, the FS contains more irregular galaxies. 

\section{Photo-$z$ performance}
\label{sec:photoz}

In this section we compute the photometric redshifts $z_{ph}$ of the galaxies in the Bright Sample (BS) and the Faint Sample (FS) generated in the previous section and analyze their photo-$z$ performance through different statistical metrics: bias, photo-$z$ precision and outlier fraction. We also apply some photo-$z$ quality cuts (\textit{odds} cuts) on the results and analyze how the performance improves. We investigate how many galaxies with poor photo-$z$ quality we need to remove in order to achieve the photo-$z$ precision requirements defined in \citet{Gaztanaga2012}.

The photo-$z$s $z_{ph}$ are obtained using the Bayesian Photometric Redshifts\footnote{\texttt{BPZ} can be found at \url{http://www.its.caltech.edu/~coe/BPZ/}.} (BPZ) template-fitting code described in~\citet{Benitez2000}. It uses Bayesian statistics to produce a posterior probability density function $p(z|m_j)$ that a galaxy is at redshift $z$ when its magnitudes in the different bands $j$ are $m_j$:
\begin{equation}
p(z|m_j) \propto \sum_t L(m_j|z,t) \, \Pi(z,t \mid m)  \, ,
\label{pz}
\end{equation}
where $L(m_j|z,t)$ is the likelihood that the galaxy has magnitudes $m_j$, if its redshift is $z$ and its spectral type $t$, and $\Pi(z,t \mid m)$ is the prior probability that the galaxy has redshift $z$ and spectral type $t$ when its magnitude in some reference band is $m$. The proportionality symbol is a reminder that $p(z|m_j)$ must be properly normalized in order to be a probability density function. The photometric redshift $z_{ph}$ of the galaxy will be taken as the position of the maximum of $p(z|m_j)$.

We have modified the {\tt BPZ} code in order to increase its efficiency when estimating photo-$z$s using a large number of narrow-band filters. Instead of estimating the
photo-$z$ for each galaxy, the calculations are done in blocks
of hundreds of galaxies using linear operations. Details will
be presented in Eriksen et al.~(in preparation).

\subsection{Templates}
\begin{figure}
\centering
\includegraphics[width=80mm]{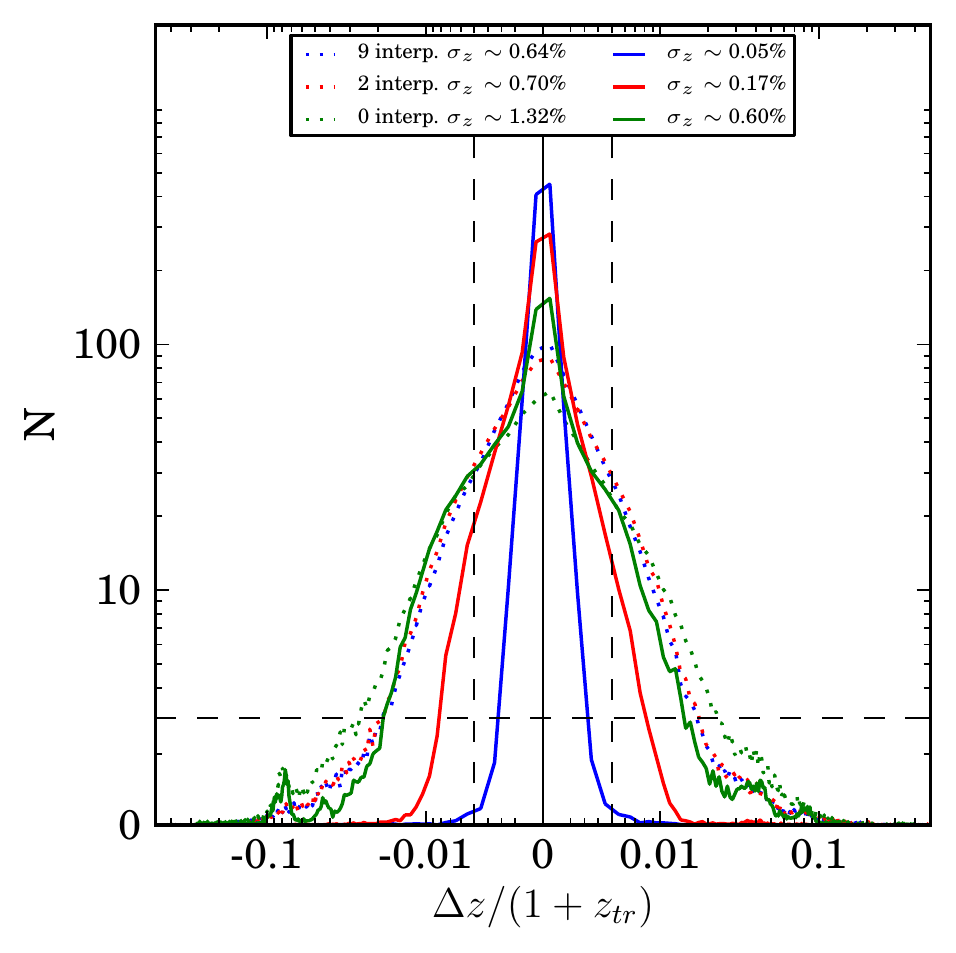}
\caption{The $\Delta z / (1+z_{tr})$ distributions for the BS using different number of interpolated templates in \texttt{BPZ}, when input magnitudes are noiseless (solid) and when they are noisy (dotted). Photo-$z$ precision $\sigma_z$ values for each case are shown on the legend. The \texttt{x}-axis scale is linear between the two dashed vertical lines and logarithmic on the sides. Similarly, the \texttt{y}-axis is linear below the horizontal dashed line and logarithmic above.}
\label{dz_hist_bright}
\end{figure}

The likelihood $L(m_j|z,t)$ is generated by comparing the observed magnitudes with the ones that are predicted through a collection of galaxy templates that span all the possible galaxy types $t$. BPZ includes its own template library; however, we use a subset of 6 templates from the same library used in the previous section for the mock catalog generation. They are highlighted in Fig.~\ref{pau_templates} and correspond to the templates with file name: \texttt{CWW\_Ell.sed}, \texttt{Ell\_09.sed}, \texttt{CWW\_Sbc.sed}, \texttt{CWW\_Scd.sed}, \texttt{CWW\_Irr.sed} and \texttt{I99\_05Gy.sed}. Additionally, we also include two interpolated templates between each consecutive pair of the six by setting the \texttt{BPZ} input parameter \texttt{INTERP=2}. This results in a total of 16 templates. However, we will see later in Fig.~\ref{dz_hist_bright} that the number of interpolated templates does not affect much the photo-$z$ performance.

\subsection{Prior}
An important point of \texttt{BPZ} is the prior probability $\Pi(z, t \mid m)$ that helps improve the photo-$z$ performance. \citet{Benitez2000} proposes the following empirical function:
\begin{eqnarray}
\Pi(z, t \mid m) &=& \Pi(t\mid m) \cdot \Pi(z \mid t, m) \nonumber \\
&\propto& f_t e^{-k_t(m-m_0)} \cdot z^{\alpha_t}\exp \left\lbrace -\left[ {z \over z_{mt}(m)} \right]^{\alpha_t} \right\rbrace \nonumber \\
\label{prior}
\end{eqnarray}
where $z_{mt}(m) = z_{0t} + k_{mt}(m-m_0)$ and $m_0$ is a reference magnitude, in our case equal to 19 in the $i$-band. Each spectral type $t$ has associated a set of five parameters $\lbrace f,k,\alpha,z_0,k_{m} \rbrace$ that determine the shape of the prior. In order to calibrate the prior $\Pi(z,t\mid m)$ and determine the value of these parameters, we construct a training sample consisting of 10000 galaxies randomly selected from the mock catalog with $i_{AB}<24$. We only need to know their observed magnitude $i_{AB}$ in our reference band, their true redshift $z_{tr}$ and their true spectral type $t_{tr}$. Originally, $t$ ranged from 0 to 66 which is the number of templates used to generate the mock catalog; however, as we did for the Luminosity Functions (LF) galaxy types in the previous section, we group all these galaxy types in three groups: $t=1$ (ellipticals), $t=2$ (spirals) and $t=3$ (irregulars), whose correspondence is 1$\rightarrow$(0-17), 2$\rightarrow$(17-55) and 2$\rightarrow$(55-66). From now on, we will use $t$ for either the 66 templates or these 3 galaxy type groups. Finally, we fit Eq.~\ref{prior} to the training sample and recover the prior parameters. We show the resulting values in Table~\ref{tab:prior}. Parameters $f_3$ and $k_3$ do not appear in the table because $\Pi(t=3 \mid m_0)$ is deduced by imposing the proper normalization.
\begin{table}
\centering
\begin{tabular}{|c||ccccc|}
\hline
$t$ & $f$ & $k$ & $\alpha$ & $z_0$ & $k_{m}$ \\ \hline
1&0.565&0.186&2.456&0.312&0.122\\
2&0.430&0.000&1.877&0.184&0.130\\
3&-&-&1.404&0.047&0.148\\
\hline
\end{tabular}
\caption{The resulting values of the prior parameters obtained by fitting (\ref{prior}) to a subset of 10000 randomly selected galaxies from the BS and FS together. The three galaxy types $t=1,2,3$ are the same that were defined in Section~\ref{sec:mock} for the Luminosity Functions (LF). $f_3$ and $k_3$ do not appear because $\Pi(t=3 \mid m_0)$ is deduced by normalization.}
\label{tab:prior}
\end{table}
\begin{figure*}
\centering
\begin{tabular}{rl}
\includegraphics[height=100mm]{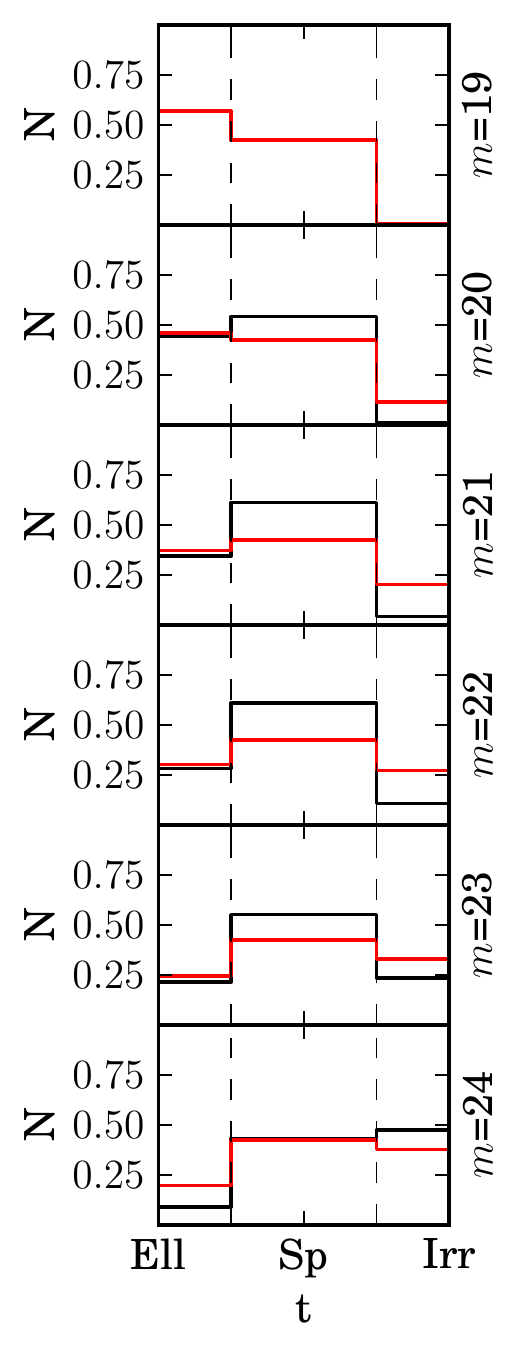} & \includegraphics[height=100mm]{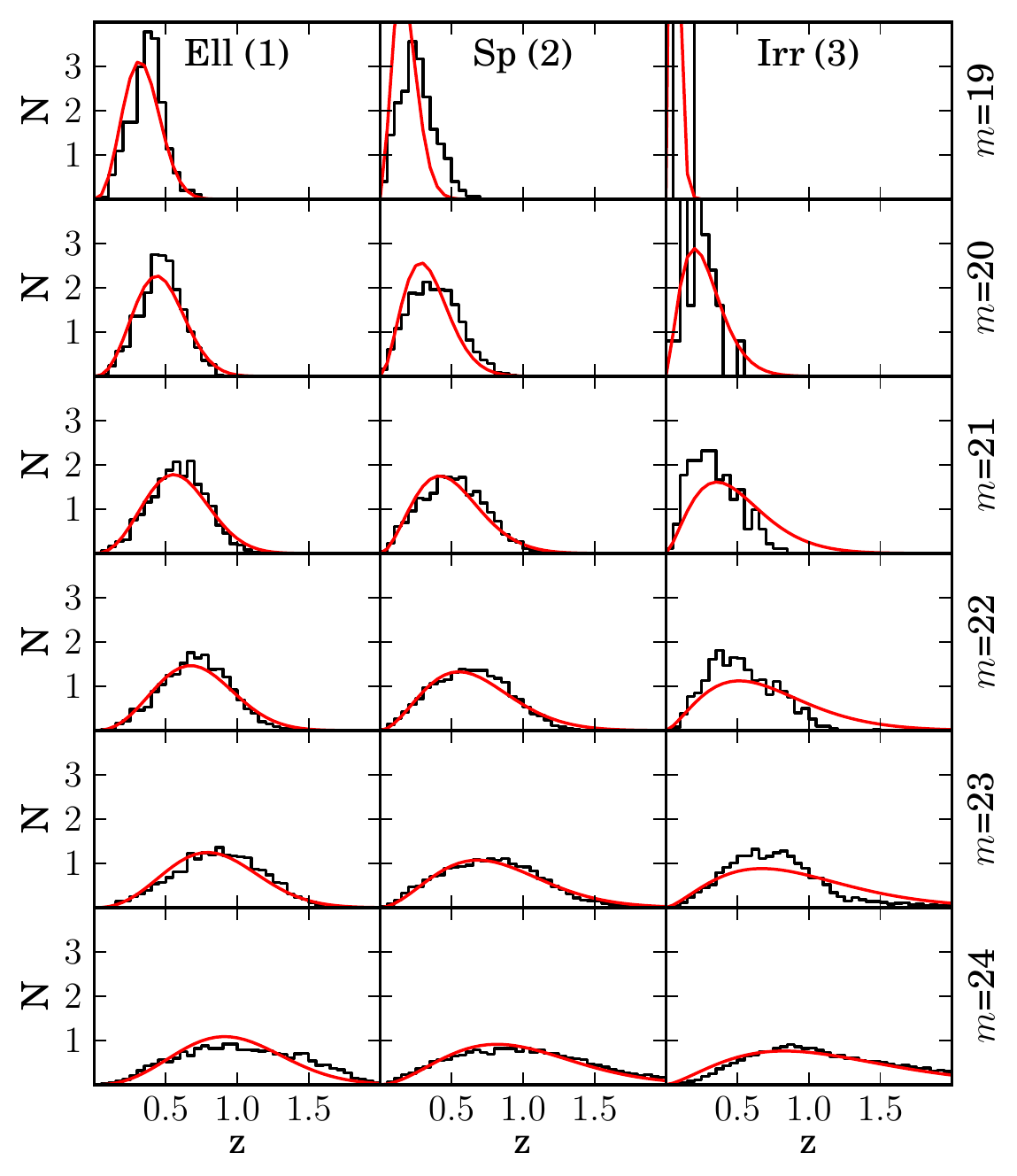} 
\end{tabular}
\caption{Left: Comparison between the fitted $\Pi(t|m)$ prior (red) of Eq.~(\ref{prior}) and the actual distribution (black). Right: The same for $\Pi(z|t,m)$. We only differentiate between 3 galaxy types $t$ (as we did for the LFs in section~\ref{sec:mock}): 1 = Elliptical, 2 = Spiral and 3 = Irregular. Rows correspond to $i_{AB}$ magnitude bins of width $\Delta m=0.2$ centered at values from $m=19$ to $24$ in steps of $1$. All curves have been normalized. Resulting prior parameters are on Table~\ref{tab:prior}.}
\label{pau_prior_plot}
\end{figure*}
In Fig.~\ref{pau_prior_plot}, we show a comparison between the fitted curve and the actual distribution of the prior. $\Pi(t|m)$ on the left and $\Pi(z|t,m)$ on the right  are shown at different magnitudes (rows) from $i_{AB}=19$, the reference magnitude, to $i_{AB}=24$. On the one hand, the fitted $\Pi(t|m)$ agrees by definition with the actual distribution at the reference magnitude (top row), since we use those values as a starting point. However, at higher magnitudes a significant mismatch appears for the spiral and irregular galaxies. This is related to the fact that the $k$ parameters, which control the migration of galaxies from one spectral type to another across magnitude, are positive definite. With this, elliptical and spiral galaxies should turn to Irregulars as the magnitude increases. However, in Fig.~\ref{pau_prior_plot} we observe that actually the spiral abundance grows slightly before starting to decrease at $m\sim21$, and this forces the fit to $k_2=0$. Consequently, elliptical galaxies migrate directly to irregulars, causing a mismatch on the pace of growth of this galaxy type abundance. On the other hand, the fit of $\Pi(z|t,m)$ is particularly good for Ell and Sp galaxies at higher magnitudes (the eight bottom-left panels on the right plot of Fig.~\ref{pau_prior_plot}), but for Irr and magnitudes close to the 19 (the reference magnitude) it is less accurate.

\begin{figure}
\centering
\includegraphics[width=80mm]{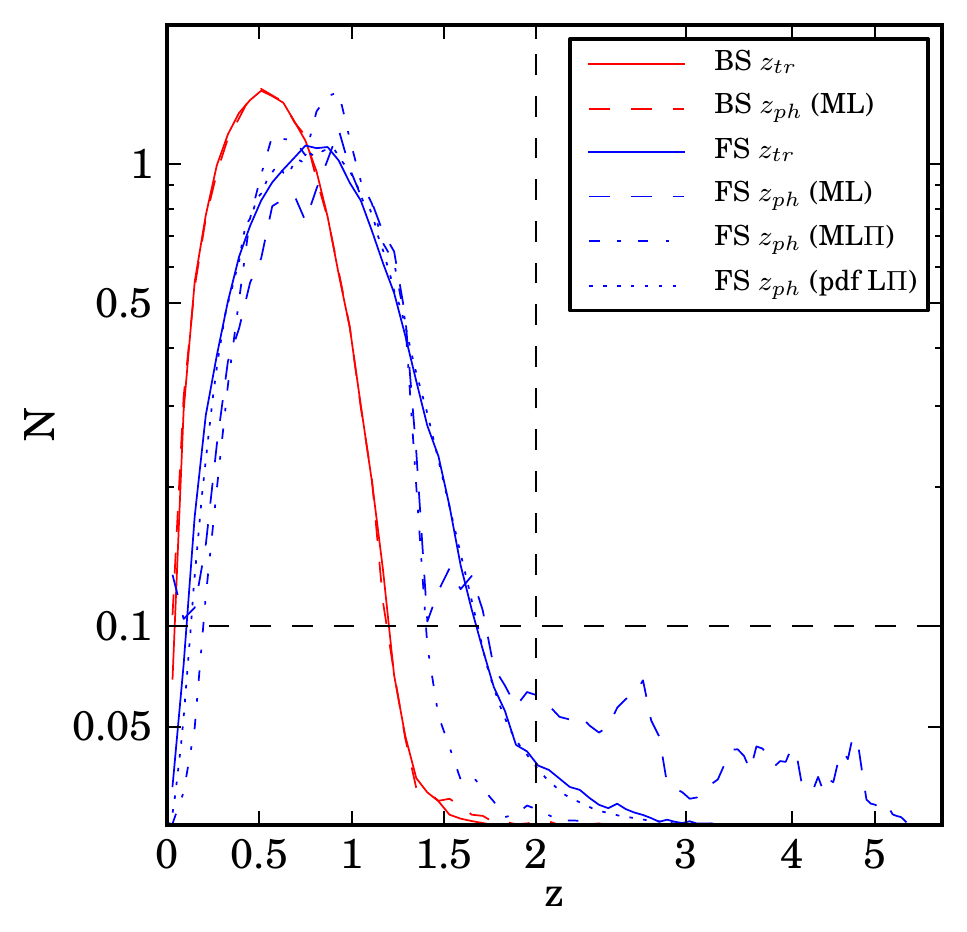}
\caption{Comparison between the true redshift (solid) and the photo-$z$ (discontinuous) distributions for the BS (red) and FS (blue). For the sake of clarity, we have set the \texttt{x}- and \texttt{y}-axis scales to be linear below the black dashed lines and logarithmic above them. All distributions have been normalized to equal area. Dashed lines correspond to the $z_{ph}$ obtained by maximizing only the likelihood $L(m_j|z,t)$ (ML) of (\ref{pz}), the dash-dotted line includes the prior (ML$\Pi$), and the dotted line corresponds to the case when all the $p(z|m_j)$ are stacked (pdf L$\Pi$).}
\label{z_hist}
\end{figure}

\begin{figure}
\centering
\includegraphics[width=80mm]{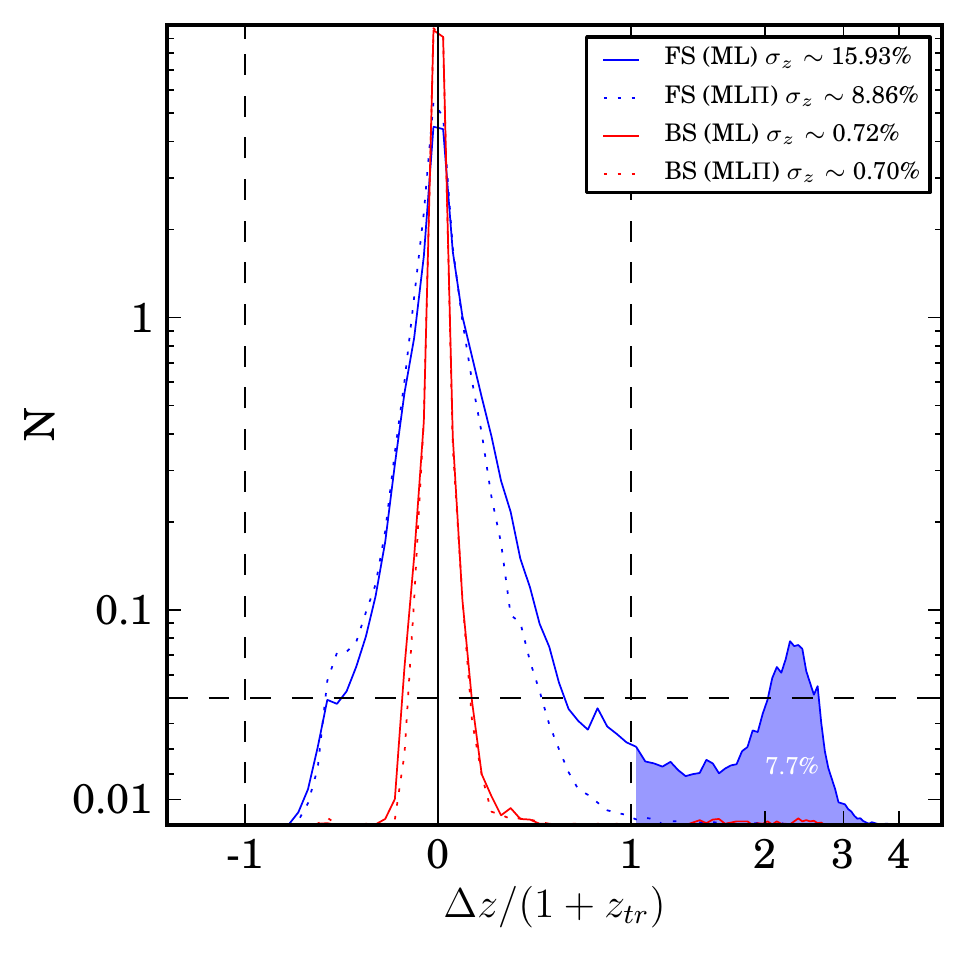}
\caption{$\Delta z / (1+z_{tr})$ distributions, where $\Delta z \equiv z_{ph} - z_{tr}$, for the BS (red) and FS (blue) with the $z_{ph}$ obtained by maximizing only the likelihood $L(m_j|z,t)$ (ML) (solid) or when also including the prior (ML$\Pi$) (dotted). Photo-$z$ precision $\sigma_z$ values, defined as half of the symmetric interval that encloses the 68\% of the distribution area around the maximum, are shown for each case in the legend. The \texttt{x}-axis scale is linear between the two dashed vertical lines and logarithmic on the sides. Similarly, the \texttt{y}-axis is linear below the horizontal dashed line and logarithmic above. The prior makes no difference on the BS, but in the FS it removes the long tail (blue region) of catastrophic outliers ($|\Delta z|/(1+z_{tr})>1$) that accounts for a $\sim 7.7\%$ of the sample. This reduces $\sigma_z$ by a factor of $\sim$1.8.}
\label{dz_hist}
\end{figure}

In Fig.~\ref{z_hist} we show a comparison between the $z_{tr}$ (solid) and $z_{ph}$ (discontinuous) distributions for the BS and the FS. For the sake of clarity, the \texttt{x}- and \texttt{y}-axis have been set to be linear below $z=2$ and $N=0.1$ respectively, and logarithmic elsewhere. Dashed lines correspond to $z_{ph}$ obtained by maximizing only the likelihood $L(m_j|z,t)$ in (\ref{pz}). This gives a $z_{ph}$ distribution in the BS very close to the actual, while in the FS a residual long tail towards much higher redshift ($z\sim5$) appears. In Fig.~\ref{dz_hist} we show the equivalent $\Delta z / (1+z_{tr})$ distributions, where $ \Delta z \equiv z_{ph} - z_{tr}$. Once again, the \texttt{x}- and \texttt{y}-axis have been set to be linear below $|\Delta z| / (1+z_{tr})=1$ and $N=0.05$ respectively, and logarithmic elsewhere. Note that the tail is also present on the right-hand side of the blue curve. If we define as \textit{catastrophic outliers} those galaxies with $|\Delta z| / (1+z_{tr}) > 1$, we find that they account for $\sim$7.7\% in the FS (the blue region under the curve) and $\sim$0.2\% in the BS. Catastrophic outliers are typically caused by degeneracies in color space, which cause confusions in the template fit and result in a much larger $|\Delta z|$. The blue-dotted line in Fig.~\ref{dz_hist} shows that when the prior is included almost all of the catastrophic outliers in the FS are removed, leaving only a small fraction of $\sim0.1\%$.
We see in Fig.~\ref{z_hist} that the $z_{ph}$ distribution after applying the prior (blue dot-dashed) decays at high redshifts faster than the $z_{tr}$ distribution. This is because we are only using the maximum of $p(z|m_j)$ for the $z_{ph}$ value. If we use the whole pdf information (blue-dotted line), the resulting $z_{ph}$ distribution is much closer to the true. Defining the photo-$z$ precision $\sigma_z$ as half of the symmetric interval that encloses the 68\% of the $\Delta z /(1+z_{tr})$ distribution area around the maximum, we find that $\sigma_z$ almost does not change in the BS (0.72\% to 0.70\%), while it improves by a factor $\sim1.8$ in the FS by going from $\sigma_z\sim16\%$ to $\sim8.86\%$ when adding the prior. In the BS sample, the likelihood function is already narrow enough, thanks to the constraining power of the narrow bands, and, hence, the prior has very limited impact on the final result.

\subsection{Performance vs. template interpolation}
At this point, we want to explore how the number of interpolated templates used in \texttt{BPZ} changes the $z_{ph}$ performance. In Fig.~\ref{dz_hist_bright} we show the $\Delta z / (1+z_{tr})$ distribution only for the BS when we use: 9 (blue), 2 (red) and 0 (green), interpolated templates. Solid lines correspond to the $z_{ph}$ obtained when the input magnitudes are noiseless (without applying Eq.~\ref{noiselesstonoisy}), while dotted lines include the noise. The $\sigma_z$ of each distribution is shown in the legend. We see that, while for noiseless magnitudes the number of interpolated templates has a significant impact on the width of the distributions and so, on their $\sigma_z$, which gets worse by a factor of $\sim3$ at each step, for noisy magnitudes these differences are smaller. In fact, going from 9 to 2 interpolated templates the differences are negligibly small and going from 2 to 0 interpolations the difference is less than a factor of 2. 
\begin{figure*}
\centering
\begin{tabular}{rl}
\includegraphics[type=pdf,ext=.pdf,read=.pdf, height=120mm]{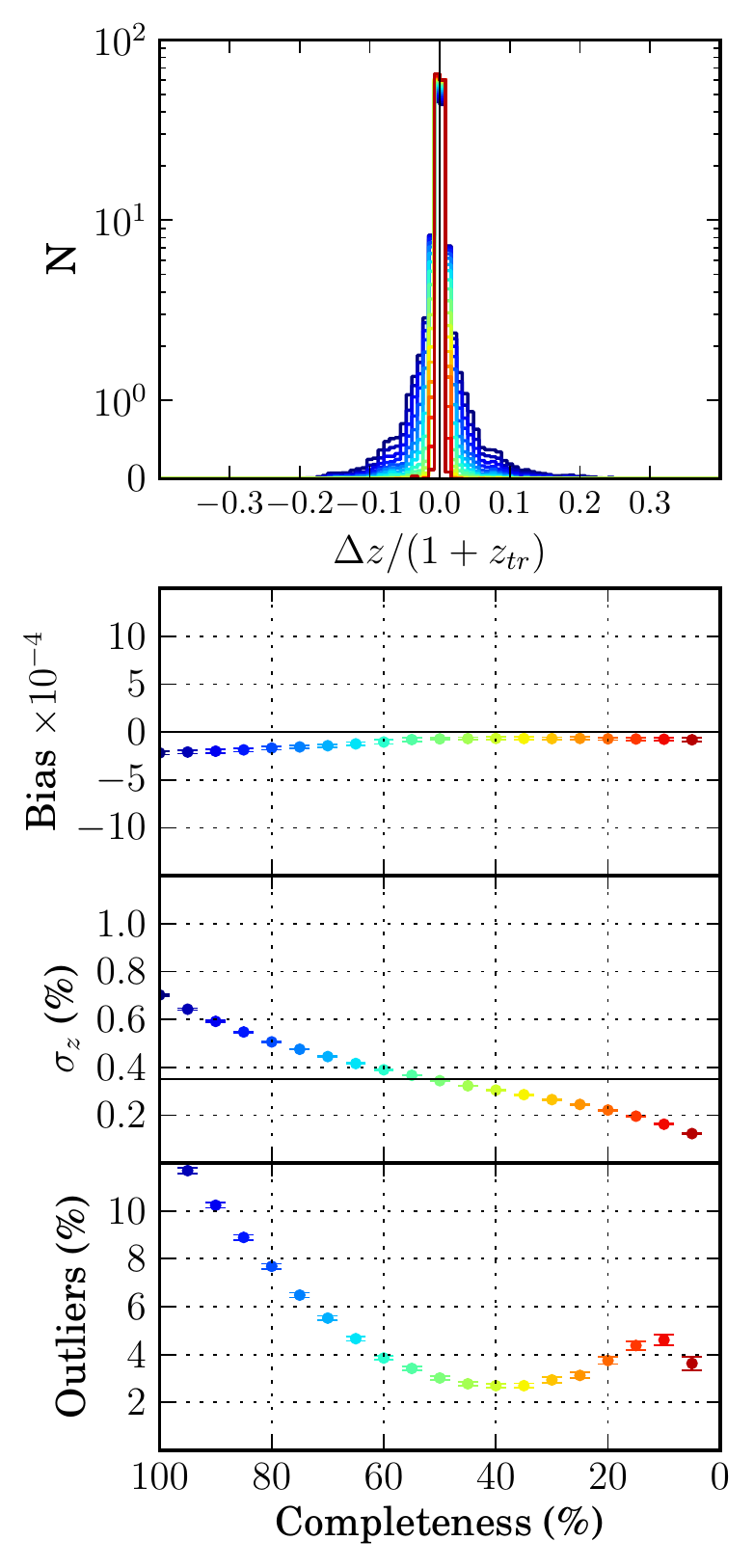} & \includegraphics[type=pdf,ext=.pdf,read=.pdf, height=120mm]{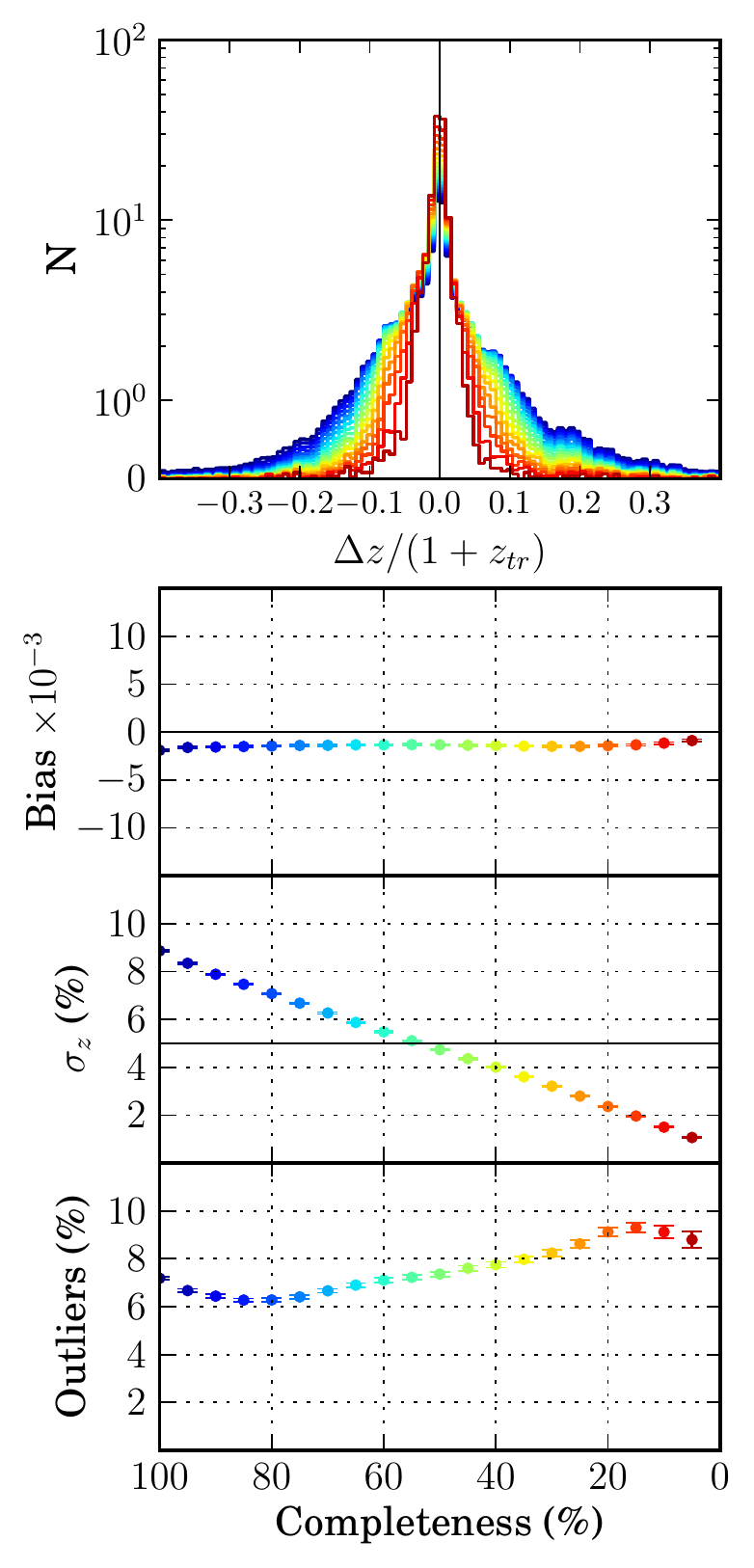}
\end{tabular}
\caption{Top: $\Delta z/(1+z_{tr})$ distributions for the BS (left) and the FS (right) at different photo-$z$ quality cuts whose completeness, shown in a color degradation, range from 100\% (bluest) to 5\% (reddest) in 5\% steps. On the bottom and by rows: the bias (median), the photo-$z$ precision ($\sigma_z$) and the $3\sigma$-outlier fraction of the distributions at the top as a function of the completeness. The colors of the points match the colors in the distributions. Note that the vertical scale in the bias and the $\sigma_z$ panels changes by an order of magnitude between the two samples. The black-dashed horizontal lines on the $\sigma_z$ panels show the photo-$z$ precision requirements defined in \citet{Gaztanaga2012}: $\sigma_z<0.35\%$ (BS) and $<5\%$ (FS).}
\label{pz_results}
\end{figure*}

\subsection{Performance vs. \textit{Odds}}
Photo-$z$ codes, besides returning a best estimate for the redshift, typically also return an indicator of the photo-$z$ quality. It can be simply an estimation of the error on $z_{ph}$, or something more complex, but the aim is the same. In \texttt{BPZ} this indicator is called \textit{odds}, and, it is defined as
\begin{equation}
odds = \int^{z_{ph}+\delta z}_{z_{ph}-\delta z}p(z|m_j)dz \, ,
\label{odds}
\end{equation}
where $\delta z$ determines the redshift interval where the integral is computed. \textit{Odds} can range from 0 to 1, and the closer to 1, the more reliable is the photo-$z$ determination, since $p(z|m_j)$ becomes sharper and most of its area is enclosed within $z_{ph}\pm \delta z$. We choose $\delta z=$ 0.0035 in the BS and 0.05 in the FS, which is close to the expected $\sigma_z$ in these samples for the PAU Survey (see the $\sigma_z$ plots in Fig.~\ref{pz_results}). A bad choice of $\delta z$ could lead to the accumulation of all \textit{odds} close to either 0 or 1. Since \textit{odds} are a proxy for the photo-$z$ quality, we should expect a correlation between the \textit{odds} and $|\Delta z|$ in the sense that higher \textit{odds} should correspond to lower $|\Delta z|$. 

At the top of Fig.~\ref{pz_results}, we show the $\Delta z/(1+z_{tr})$ distributions in a color degradation for subsets of the BS (left) and the FS (right) with increasingly higher cuts on the \textit{odds} parameter. In fact, the exact \textit{odds} values are quite arbitrary, since they depend on the size of $\delta z$. Therefore, we have translated these \textit{odds} cuts into the fraction of the galaxy sample remaining after a certain cut has been applied, in such a way that the bluest curve corresponds to 100\% completeness while the reddest corresponds to 5\% completeness, with 5\% steps. We can clearly see how the harder are the \textit{odds} cuts, the narrower and peaky become the distributions in both samples. On the bottom plots of the same figure we show how some statistical metrics of these distributions (the bias (median), the photo-$z$ precision ($\sigma_z$) and the $3\sigma$-outlier fraction) depend on each \textit{odds} cut of completeness given in the \texttt{x}-axis. The $3\sigma$-outlier fraction is defined as the fraction of galaxies with $|\Delta z|/(1+z_{tr})>3\sigma_z$. For the sake of clarity, each point has been colored as its correspondent distribution. Errors are computed by bootstrap \citep{efron79} for the bias and $\sigma_z$, and, by computing the $\sigma_{68}$ of a binomial distribution with mean $n_{outlier} / N$ for the outlier fraction. As we expected, $\sigma_z$ decreases as the \textit{odds} cuts get more stringent. In the BS (left), it goes from $\sim$0.7\% at 100\% of completeness to $\sim$0.1\% at 5\%, and in the FS (right), from $\sim$9\% to $\sim$1\%. The photo-$z$ precision requirements, as defined in \citet{Gaztanaga2012}, are $\sigma_z < 0.35\%$ in the BS and $\sigma_z < 5\%$ in the FS. They are fulfilled when $\sim$50\% of each catalog is removed. We find a very small bias of a few percent of $\sigma_z$ in both samples towards negative $\Delta z$ values. It practically vanishes when high \textit{odds} cuts are applied. The $3\sigma$-outlier fraction in the BS starts at $\sim$13\%, drops to $\sim$3\% at $\sim$40\% completeness and then, starts increasing again up to $\sim$4.5\% at $\sim$10\% completeness. Therefore, we deduce that the gain on $\sigma_z$ with the \textit{odds} cut occurs basically through the cleaning of outliers. However, in the FS, even if $\sigma_z$ decreases with the \textit{odds} cuts, the outlier fraction increases from $\sim$7\% to $\sim$9\% at $\sim$15\% completeness. 

\begin{figure*}
\includegraphics[type=jpg,ext=.jpg,read=.jpg, width=130mm]{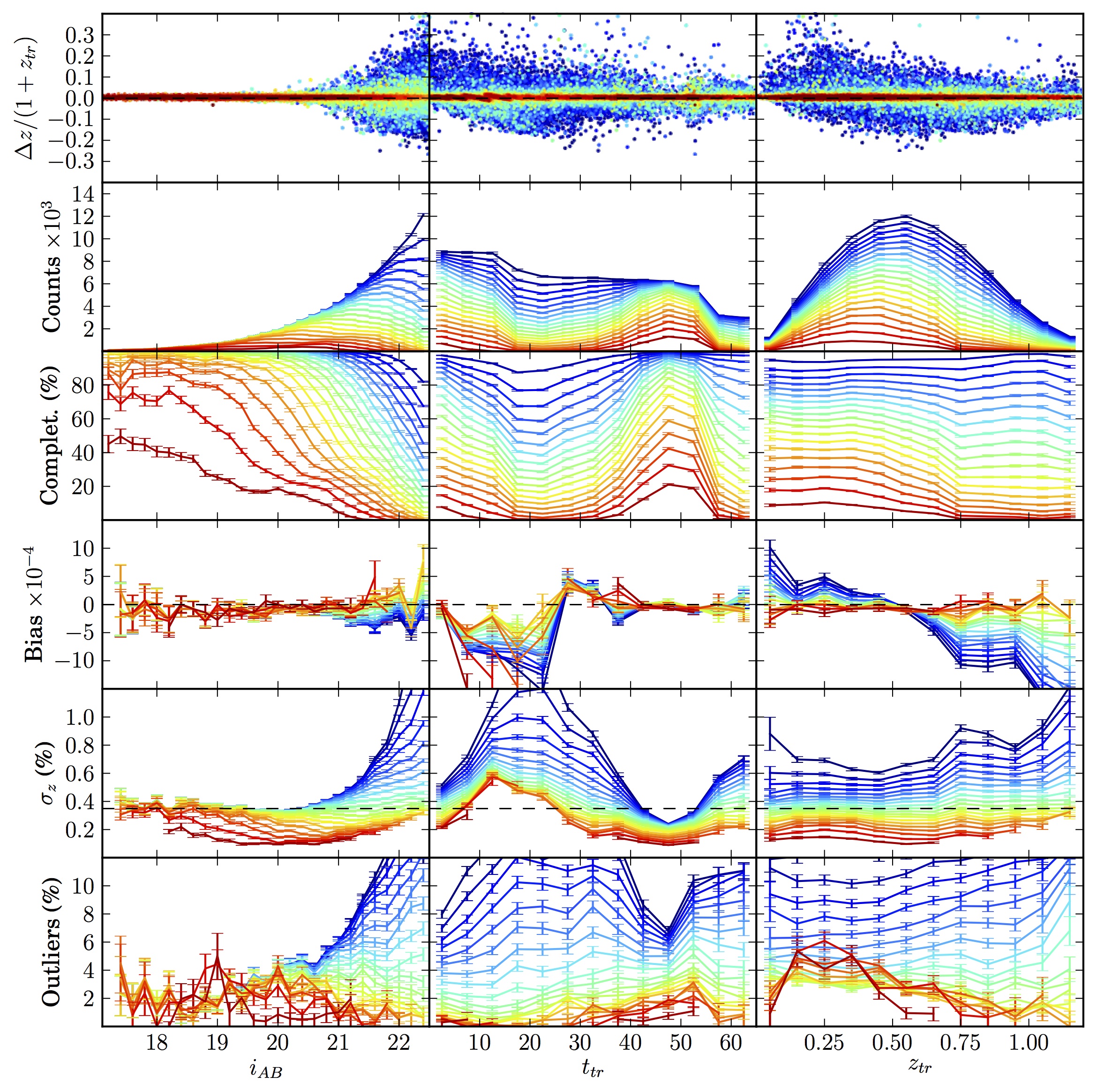}
\caption{Statistics showing the PAU-BS photo-$z$ performance as a function of the observed $i_{AB}$ magnitude (left), true galaxy type $t_{tr}$ (center) and true redshift $z_{tr}$ (right). In the first row we show the scatter $\Delta z/(1+z_{tr})$. Then, in a descending order of rows, we show the number of galaxies, the completeness, the bias (median), the photo-$z$ precision ($\sigma_z$) and the $3\sigma$-outlier fraction for all the \textit{odds} cuts shown in Fig.~\ref{pz_results} (same color). Also as in Fig.~\ref{pz_results}, the black-solid horizontal lines on the $\sigma_z$ panels show the photo-$z$ precision requirement defined in \citet{Gaztanaga2012}.}
\label{bs_pz_results}
\end{figure*}

\begin{figure*}
\includegraphics[type=jpg,ext=.jpg,read=.jpg, width=130mm]{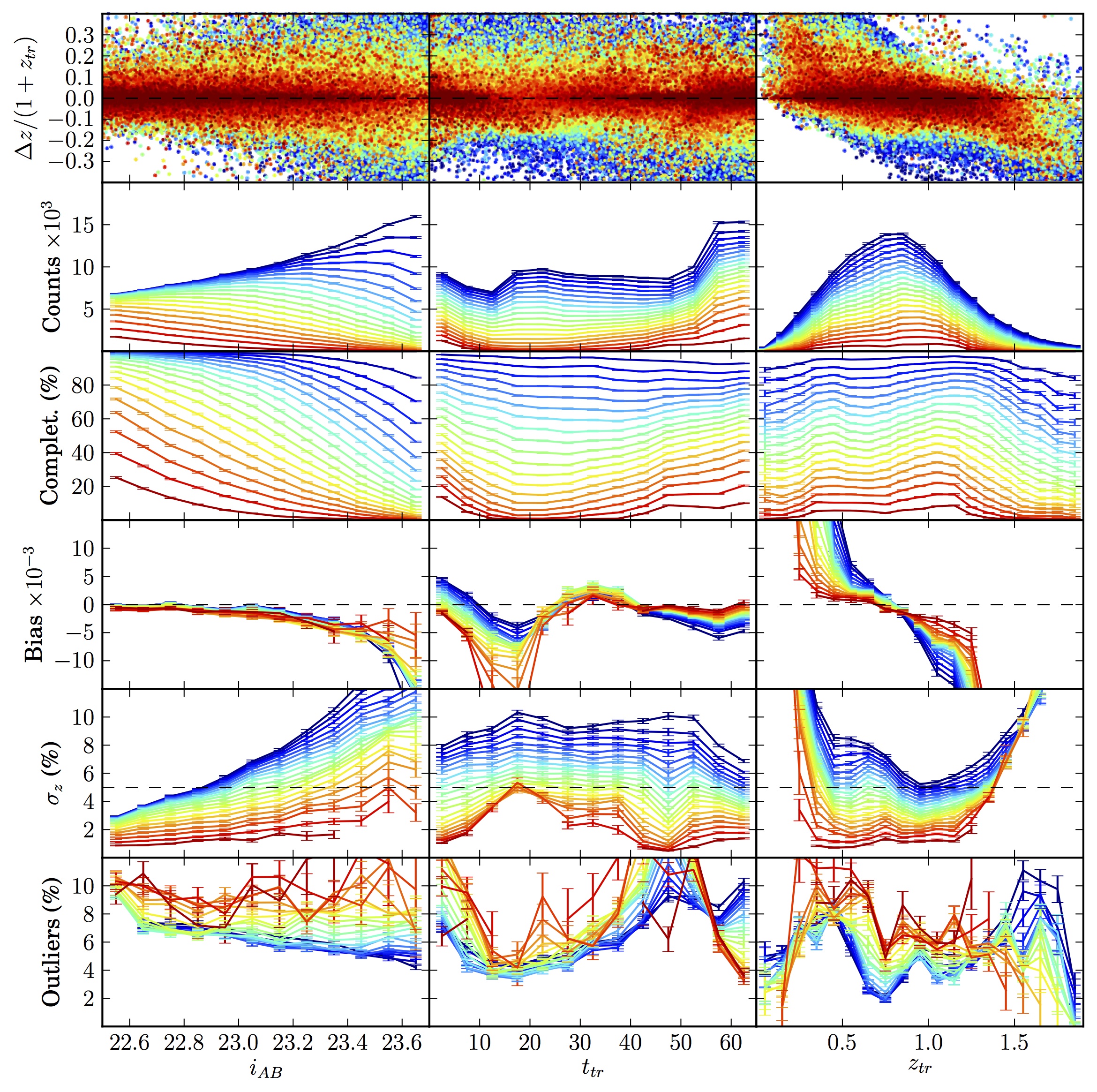}
\caption{Statistics showing the PAU-FS photo-$z$ performance in the same layout as in Fig.~\ref{bs_pz_results}.}
\label{fs_pz_results}
\end{figure*}

On the last three rows of plots in Figs.~\ref{bs_pz_results} (BS) and~\ref{fs_pz_results} (FS), we show how these statistical metrics depend on the observed magnitude $i_{AB}$ (left), the true spectral type $t_{tr}$ (center) and the true redshift $z_{tr}$ (right), after each photo-$z$ quality cut shown in Fig.~\ref{pz_results} in the same color. In the first three rows and by order, we also show the scatter plot $\Delta z / (1+z_{tr})$, the number of galaxies and the completeness after the same photo-$z$ quality cuts with respect to the same variables ($i_{AB}$, $t_{tr}$,$z_{tr}$) on the \texttt{x}-axis.

In the BS (Fig.~\ref{bs_pz_results}), we can see that the low photo-$z$ quality galaxies (blue points in the scatter plot) are mostly faint galaxies with $i_{AB}>21$, the magnitude where the noise coming from the sky brightness plus the CCD read-out starts to be comparable to the Poisson noise in signal. This is reflected in Fig.~\ref{err_m} as a turning point on the slope of the $\sigma_m$ vs. $m_{AB}$ scatter. In fact, these galaxies represent most of the outliers and the principal source of bias seen in Fig.~\ref{pz_results}. As the \textit{odds} cuts are applied, these bad photo-$z$ faint galaxies are removed. The \textit{odds} cut removes the bias, reduces $\sigma_z$ from $\sim$2.2\% to $\sim$0.35\% and the outlier fraction from $>$10\% to $\sim$1\% at magnitudes close to the limit $i_{AB}=22.5$. Looking at the scatter plots of the next two columns in Fig.~\ref{bs_pz_results}, we realize that these low-\textit{odds} galaxies at faint magnitude are spread out over the whole $t_{tr}$ and $z_{tr}$ ranges. Moreover, after the hardest \textit{odds} cut, only galaxies of types $t\sim0$ (elliptical) and $t\sim50$ (irregular) survive and the mean of $z_{tr}$ is shifted from $\sim$0.6 to $\sim$0.4. The worst bias, $\sigma_z$ and outlier fraction are obtained for Spiral galaxies ($t\sim$10-30). The \textit{odds} cuts mitigate these results, but even after applying them, spiral galaxies still have the worst bias and $\sigma_z$. The worst bias is located at low and high $z_{tr}$, with opposite sign and it is largely reduced with the \textit{odds} cuts. The value of $\sigma_z$ gets flatter over all $z_{tr}$ as the \textit{odds} cuts are harder.

Regarding the $z_{ph}$ precision requirement $\sigma_z < 0.35\%$ (black solid horizontal line), we find that, when no \textit{odds} cut is applied, it is achieved only for galaxies with $i_{AB}<21$ and galaxy type around $t\sim50$ (irregulars). However, it is not fulfilled at any $z_{tr}$. Once we apply a 50\% completeness \textit{odds} cut, which gives an overall $\sigma_z$ equal to the requirement, as we saw in Fig.~\ref{pz_results}, the requirement is fulfilled in all the $i_{AB}$ and $z_{tr}$ ranges. Only for spiral galaxies the requirement is not fulfilled even after the hardest \textit{odds} cut. Originally, in \citet{Benitez2009}, it was assumed that elliptical galaxies (or rather Luminous Red Galaxies) yield the best photo-$z$ precision in the PAU Survey, with the narrow bands tracking the $\sim$4000\AA \ break spectral feature (Fig.~\ref{pau_templates}). This is partially true, since we actually see that elliptical galaxies give better $\sigma_z$ than spirals, but our analysis shows that in fact irregulars with $t\sim50$ give the best photo-$z$ performance. Before any \textit{odds} cut, their photo-$z$ precision is almost twice better than the requirement. Probably this is due to the fact that, in contrast to elliptical galaxies were a single spectral feature is tracked, irregulars have the two emission lines [OII] $\sim$3737\AA \ and [OIII] $\sim$5000\AA \ (Fig.~\ref{pau_templates}) with intrinsic widths narrower than the width of the NB filters.

In the FS (Fig.~\ref{fs_pz_results}), we see that the scatter of $\Delta z/(1+z_{tr})$ is much larger, as expected from the wider histograms in Fig.~\ref{pz_results}. However, a wider but still tight core close to $\Delta z = 0$ with high photo-$z$ quality (red points) remains. We recognize behaviors similar to those in the BS in most aspects of the $z_{ph}$ performance, although they are substantially larger. For example, the highest magnitude as well as lowest and highest $z_{tr}$ galaxies are the most biased. The \textit{odds} cuts also mitigate this bias, but a residual bias of opposite sign persists at the extremes of $z_{tr}$. Spiral galaxies ($t\sim10$-$30$) are also the ones with the highest bias, and the \textit{odds} cuts even aggravates this. We also see that elliptical ($t\sim0$) and irregular ($t>50$) galaxies are initially biased, but, in contrast to spirals, the \textit{odds} cuts help to reduce the bias. Unlike in the BS, $\sigma_z$ increases along all the magnitude range since at those magnitudes the noise from the sky brightness dominates over the signal (Fig.~\ref{err_m}). However, we see that the slope of the $\sigma_z$ increase is smaller the harder the \textit{odds} cuts. This is because the gain in photo-$z$ precision is at the expense of keeping only brighter galaxies each time. The mean $i_{AB}$ magnitude goes from $\sim$23.2 to $\sim$22.8 with the \textit{odds} cuts, and the shift in the $z_{tr}$ mean is from $\sim$0.86 to $\sim$0.81. Unlike in the BS, we see that the best $\sigma_z$ is obtained at the extremal spectral types: $t\sim0$ (elliptical) and $t\sim66$ (irregular). However, once the hardest \textit{odds} cuts are applied, irregular galaxies with $t\sim$50 are again the once with the best $\sigma_z$. In fact, the hardest \textit{odds} cuts also remove all spiral galaxies. Note that the large bias seen at the extremes of $z_{tr}$ make $\sigma_z$ take values much larger at these redshifts. The photo-$z$ precision requirement, $\sigma_z<5\%$, is fulfilled when the \textit{odds} cut of 50\% completeness is applied up to magnitude $\sim23.1$, for all galaxy types except spirals, and at the $z_{tr}$ interval from $\sim0.4$ to $\sim1.3$. As we already saw in Fig.~\ref{pz_results}, the $3\sigma$-outlier fraction grows with the \textit{odds} cuts. In general, its values are higher where $\sigma_z$ is lower, since the outliers criterion becomes more stringent.

\subsection{Narrow bands vs. Broad bands}
\begin{figure}
\centering
\includegraphics[height=80mm]{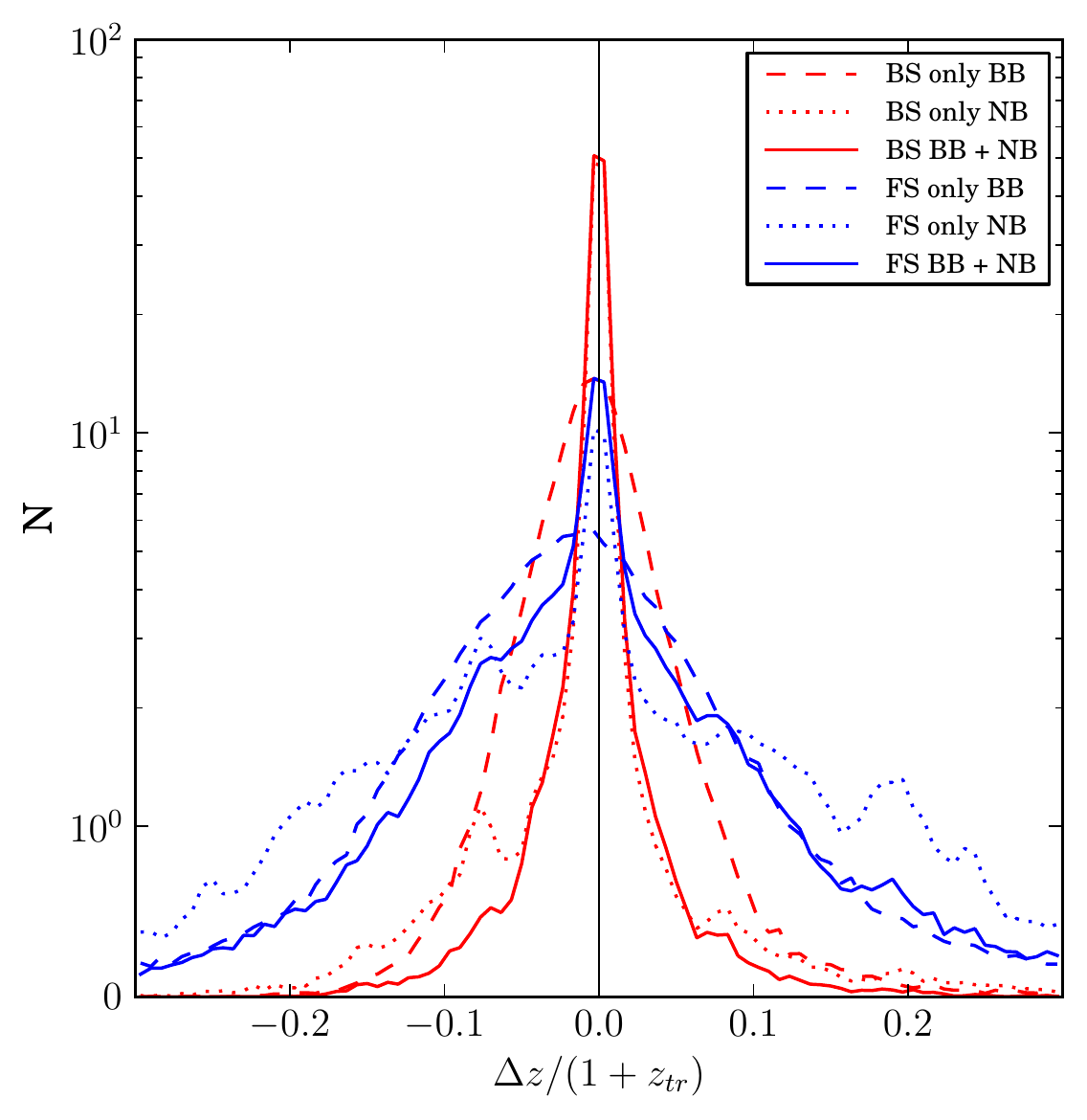}
\caption{Normalized $\Delta z/(1+z_{tr})$ distributions for the BS (red) and the FS (blue) using only BB (dashed), only NB (dotted) or both together BB+NB (solid). Bias, $\sigma_z$ and 3$\sigma$-outlier fraction of each distribution are shown in Table~\ref{tab:usefulness_NB}.}
\label{Dz_pau_only_BB}
\end{figure}

We want to quantify the improvement that the NB bring to the photo-$z$ performance. For this purpose, we run \texttt{BPZ} on the BS and the FS using the NB and the BB separately. Then, in Fig.~\ref{Dz_pau_only_BB} and Table~\ref{tab:usefulness_NB}, we compare results between these runs and also with the original ones when the BB and the NB are used together (BB+NB). Figure~\ref{Dz_pau_only_BB} shows normalized $\Delta z/(1+z_{tr})$ distributions for the BS (red) and the FS (blue) using only the BB (dashed), only the NB (dotted) and both together BB+NB (solid). We see that the resulting distributions when using only BB (dashed) show overall shapes close to Gaussian with perhaps larger tails on both sides. However, when the NB are also included (solid) the peaks of the distributions become clearly sharper. This is more noticeable in the BS than in the FS, because the non-observed condition ($\sigma_m<0.5$) defined in Section~\ref{sec:mock} implies that most of the NB are not used in the photo-$z$ determination for the FS. Table~\ref{tab:usefulness_NB} shows bias (median), $\sigma_z$ ($\sigma_{68}$) and $3\sigma$-outlier fraction of each distribution. $\sigma_z$ in the BS is reduced $\sim$4.8 times going from $\sim$3.34\% to $\sim$0.7\% when the NB are included, while the improvement is much less significant in the FS. We also see that bias is reduced by an order of magnitude when the NB are included in both samples. On the contrary, the outlier fraction increases, but as mentioned, this is due to the fact that improvements on $\sigma_z$ penalize the outlier fraction.  On the other hand, we see that using NB alone slightly degrades all metrics in the BS and the FS, except the outlier fraction in the FS which is improved for the same reason. In fact, $\sigma_z$ in the FS gets almost twice worse than when only using BB or BB+NB. It seems that in the FS NB by themselves only help the bias, while if they are used together with the BB, the improvement also extends to $\sigma_z$. These results are in qualitative agreement with previous findings using photometric systems mixing BB and NB, such as those in~\citet{Wolf2001a}.

\begin{table}
\centering
\begin{tabular}{c|ccc|c|ccc|}
\cline{2-4}
 & \multicolumn{3}{|c|}{Bright Sample} \\
\cline{2-4}
 & BB & NB & BB + NB \\
\cline{1-4}
\multicolumn{1}{|c|}{Bias$\times10^{-4}$} & -31.64 & -3.31 & -2.18 \\
\multicolumn{1}{|c|}{$\sigma_z$(\%)} & 3.34 & 0.83 & 0.70 \\
\multicolumn{1}{|c|}{Outliers(\%)} & 4.41 & 18.19 & 13.28 \\
\cline{1-4}
 & \multicolumn{3}{|c|}{Faint Sample} \\
\cline{2-4}
 & BB & NB & BB + NB \\
\cline{1-4}
\multicolumn{1}{|c|}{Bias$\times10^{-4}$} & -152.66 & -41.19 & -19.01 \\
\multicolumn{1}{|c|}{$\sigma_z$(\%)} & 9.38 & 16.17 & 8.86 \\
\multicolumn{1}{|c|}{Outliers(\%)} & 6.79 & 4.90 & 7.18 \\
\cline{1-4}
\end{tabular}
\caption{Bias (median), $\sigma_z$ ($\sigma_{68}$) and $3\sigma$-outlier fraction when using only BB, only NB or both together BB+NB for the bright and faint samples.}
\label{tab:usefulness_NB}
\end{table}

\subsection{Impact of the photo-$z$s on the clustering}
\begin{figure}
\centering
\includegraphics[width=80mm]{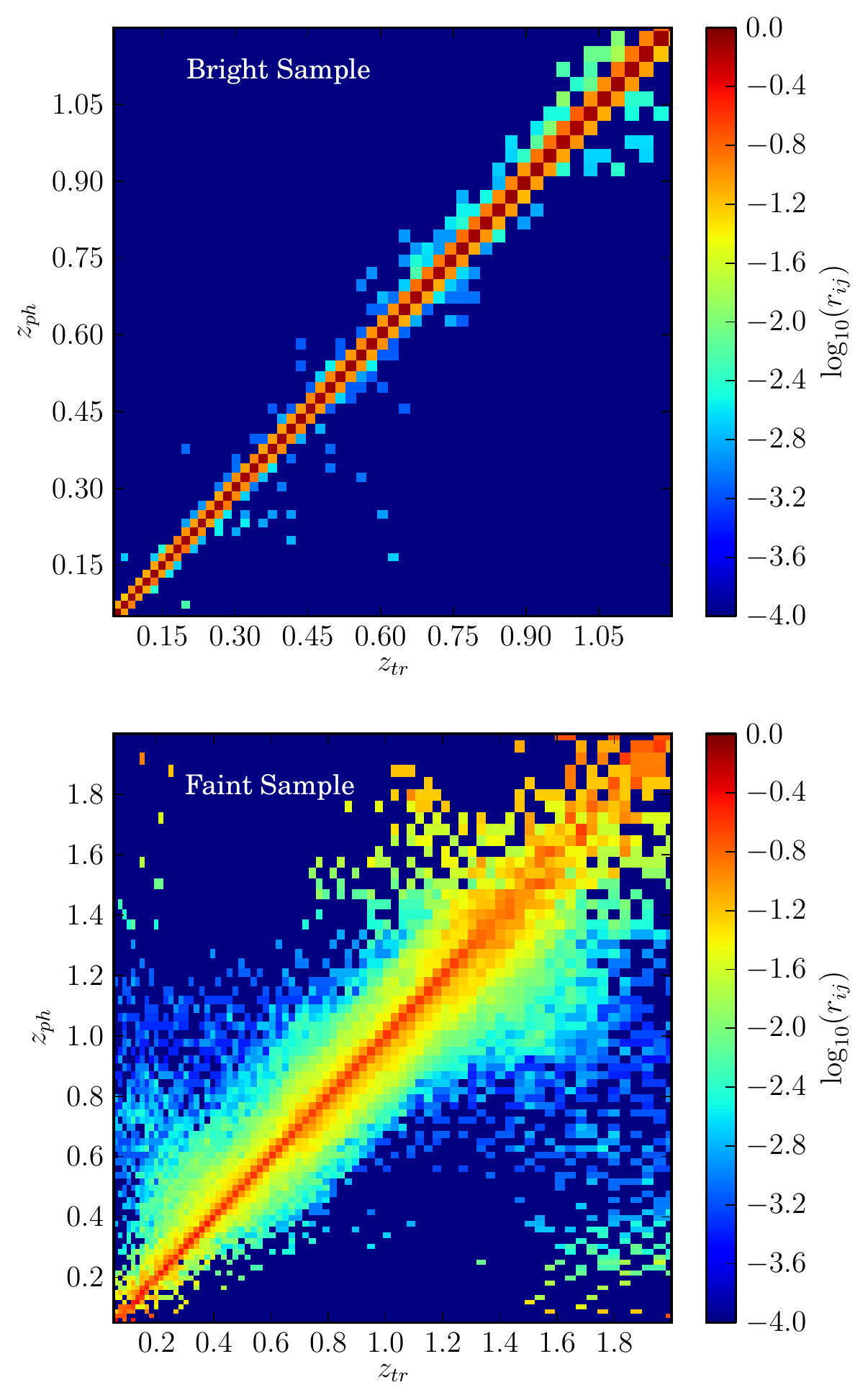}
\caption{The resulting migration matrices $r_{ij}$ from the PAU photo-$z$ simulations after applying the photo-$z$ quality cut that leaves 50\% completeness. The top plot corresponds to the BS and the bottom plot to the FS. For a higher contrast and clarity we plot the logarithm of the matrix values. These matrices give the probability that a galaxy observed at the photo-$z$ bin $i$, will be actually at the true redshift bin $j$. The bin widths are $0.014(1+z)$, four times the photo-$z$ precision expected in the BS.}
\label{plot:rij}
\end{figure}

We want to study the impact of the PAU photo-$z$ performance on the measurements of angular clustering. In \citet{Gaztanaga2012} it is shown that galaxy cross-correlation measurements $\bar{\omega}_{ij}$ between two photo-$z$ bins $i$ and $j$ are related to their cross-correlation between real redshift bins $\omega_{ij}$ as
\begin{equation}
\bar{\omega}^{A\times B}_{ij} = \sum_{kl} r^A_{ik} \omega^{A \times B}_{kl} r^B_{jl} = r_A \cdot \omega^{A \times B} \cdot  r^T_B,
\label{eq:wijbar}
\end{equation}
where $r_{ij}$ is called the migration matrix and gives the probability that a galaxy observed at the photo-$z$ bin $i$ will be actually at the true redshift bin $j$, while $A$ and $B$ denote different galaxy samples. We compute the migration matrices from the PAU photo-$z$ simulations and show them in Fig.~\ref{plot:rij} for the BS (top) and the FS (bottom) in photo-$z$ bins of width $0.014(1+z)$, which is four times the photo-$z$ precision $\sigma_z$ in the BS once the photo-$z$ quality cut that leaves a 50\% completeness is applied. Note that the matrices are normalized row-wise by definition. The FS migration matrix has quite a  thick diagonal, i.e. around $z_{tr} \simeq 1$ the width is $\Delta z \simeq 0.1$ for $\simeq 10\%$ probabilities and $\Delta z \simeq 0.4$ for $\simeq 1\%$. There are also outliers going from very large true redshifts 
$z_{tr} \simeq 1.8$ to lower photo-$z$ redshifts and some from low true redshifts up to  $z_{ph} \simeq 1$. The  BS has a 
thinner diagonal with $\Delta z < 0.04$ at $\simeq 1\%$ probability and with fewer outliers.
These values are of course in agreement with previous results in Figs.~\ref{dz_hist} and \ref{pz_results}.

\begin{figure*}
\centering
\includegraphics[width=170mm]{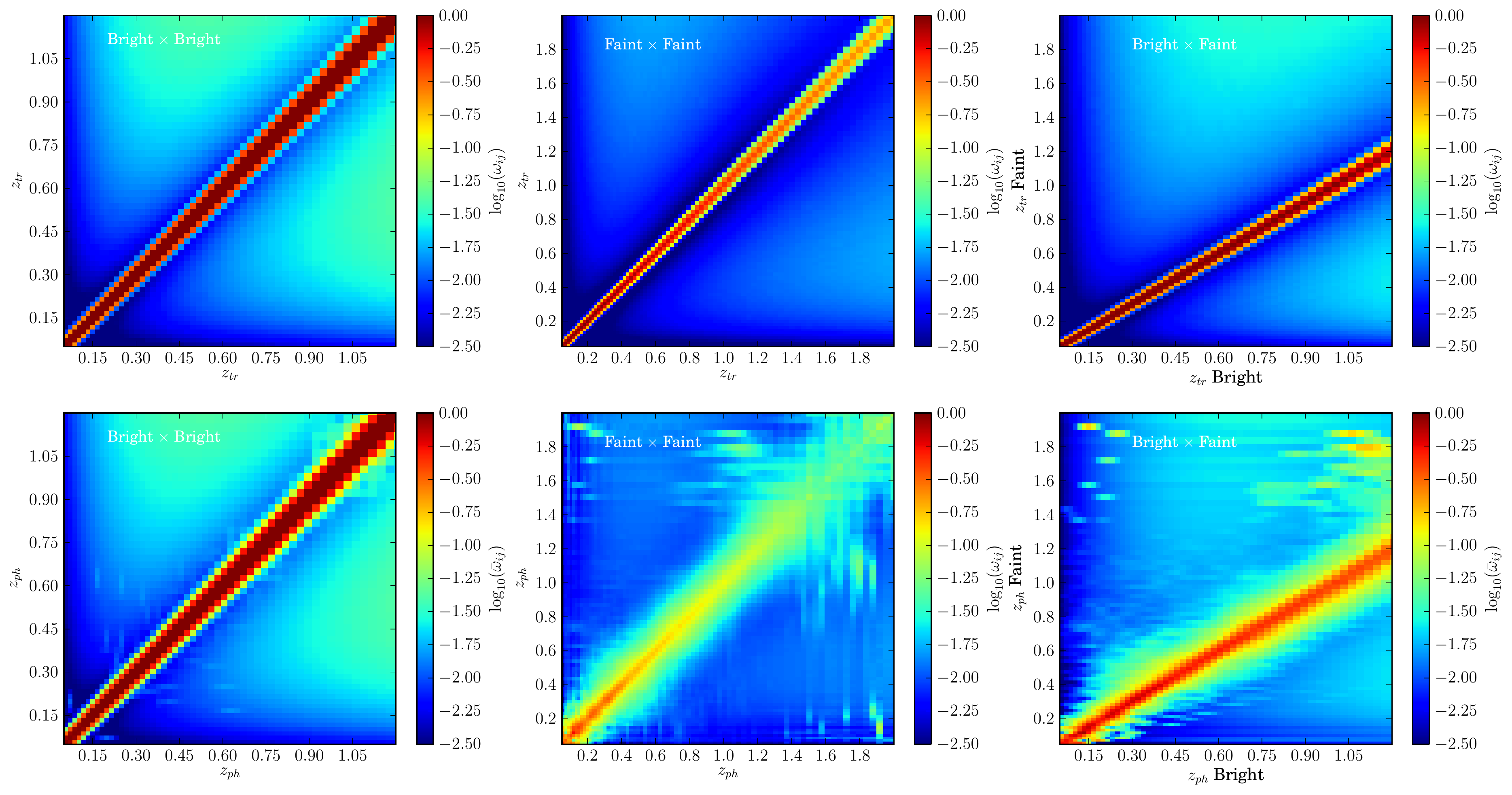}
\caption{Top panels show the angular auto (diagonal) and cross- (off-diagonal) correlations $\omega_{ij}$ at 1~arcmin between the true redshift bins $i$ and $j$ of the BS (left), the FS (middle) and the crossing of both of them (right). Bottom plots show the same correlations but measured with photo-$z$ bins, which can be computed by using Eq. (\ref{eq:wijbar}) with the migration matrices in Fig.~\ref{plot:rij}. The bin widths are $0.014(1+z)$, four times the photo-$z$ precision expected in the Bright Sample.}
\label{plot:wij}
\end{figure*}

To illustrate how these matrices affect angular clustering measurements, we use a simple model to
predict $\omega_{ij}$ at an arbitrary reference angular scale, $\theta$, of 1~arcmin. We include intrinsic
galaxy clustering and weak lensing magnification:
\begin{eqnarray}
w_{ij} &=& w_{G_iG_j}+ w_{G_i\mu_j} + w_{\mu_i  G_j}+ w_{\mu_i\mu_j} \\ \nonumber
w_{G_iG_j} &=&  b_ib_j \int dz_1dz_2  \, \, \phi_{G_i}(z_1) \, \phi_{G_j}(z_2)  \, \, \xi(r_{12}) \\\nonumber
 w_{G_i\mu_j} & =&   b_i \alpha_j \int dz_1 dz_2 \, \,\phi_{G_i}(z_1) \, p_{\mu_j}(z_2) \, \, \xi(r_{12}) \\\nonumber
w_{\mu_i\mu_j} & =&   \alpha_i \alpha_j \int dz_1 dz_2 \,  \, p_{\mu_i}(z_1) \, p_{\mu_j}(z_2) \, \, \xi(r_{12})
\label{eq:all4}
\end{eqnarray}
where $\xi(r_{12})$ is the non-linear matter 2-point correlation 
between the positions of two galaxies separated in 3D space by $r_{12}=r_2-r_1$, 
where the angular separation between $r_1$ and $r_2$ is fixed to be the reference angle (i.e. $\theta=1$ arcmin),
while the radial separation is integrated out via $z_1$ and $z_2$. We have that
$\phi_{G_i}(z)$ is a top-hat distribution for galaxies in the  redshift bin $i$  and $p_{\mu_j}(z)$ is the efficiency of weak
lensing effect for lenses at $z$ and sources at $z_j$ (following the notation in \citet{Gaztanaga2012}). 
The coefficient $b_i$ is the effective galaxy bias at $z_i$  and 
 $\alpha_j\equiv 2.5s_j-1$ is the amplitude of the weak lensing magnification effect (with $s_j$ the slope
of the galaxy number counts at the flux limit of the sample at $z_j$). 
The first equation above has 4 terms corresponding to galaxy-galaxy
(intrinsic clustering), galaxy-magnification, magnification-galaxy  and 
magnification-magnification correlation. In our test, we use $\alpha_i=1$ and $b_i=1$
to generate the starting point for $\omega_{ij}$. We then apply
the following transformation:
\begin{eqnarray}
\omega^{B\times B}_{ij} \rightarrow b_{B}(z_i)b_{B}(z_j)\omega_{ij} \\
\omega^{F\times F}_{ij} \rightarrow b_{F}(z_i)b_{F}(z_j)\omega_{ij} \\
\omega^{B\times F}_{ij} \rightarrow b_{B}(z_i)b_{F}(z_j)\omega_{ij}
\end{eqnarray}
depending on which galaxy samples we are cross-correlating (FS or BS), where
\begin{eqnarray}
b_{B}(z_i) &=& 2 + 2 (z_i - 0.5) \\
b_{F}(z_i) &=& 1.2 + 0.4 (z_i - 0.5)
\end{eqnarray}
are the biases in the BS and FS respectively in the photo-$z$ bin $i$ (at mean redshift $z_i$). This corresponds
to using linear bias for the intrinsic  correlation (as in \citet{Gaztanaga2012}) and some particular evolving
slope ($s_i \sim 1$) for the magnification cross-correlations.
 In Fig.~\ref{plot:wij} we show $\omega^{B\times B}_{ij}$ (left), $\omega^{F\times F}_{ij}$ (middle) and $\omega^{B\times F}_{ij}$ (right) before (top) and after (bottom) being transformed by the migration matrices $r_{ij}$ in Fig.~\ref{plot:rij} through Eq. (\ref{eq:wijbar}). Note that for the $B\times B$ and $F \times F$ cases the correlations matrices are symmetrical. Also note that the redshift ranges in both samples are different, so that the $B \times F$ correlation matrix is not squared.

 In the top panels, we can see the
intrinsic galaxy-galaxy clustering in the diagonal of the matrix, which has an amplitude of order unity and
decreases rapidly to zero for separated redshift bins. 
The galaxy-magnification correlation appears as a diffused off-diagonal cloud with an 
amplitude $<0.05$ (clear colors). The magnification-magnification contribution is negligible.
In the bottom panel, we see the effect of the photo-$z$ migration. The diagonal
(auto-correlations) becomes thicker and diluted. The off-diagonal galaxy-magnification cloud becomes
 more diffused, specially for the FS. The cross-correlation $F \times B$ produces results that are intermediate
between $B \times B$ and $F \times F$.

We can invert the migration matrices to go from the bottom
panels (which are the observations $\bar{\omega}^{A\times B}$) to the top panels (i.e. true correlations
$\omega^{A \times B}$) by inverting Eq. (\ref{eq:wijbar}):

\begin{equation}
 \omega^{A \times B} = r_A^{-1} \cdot \bar{\omega}^{A\times B}  \cdot  (r^T_B)^{-1} 
\label{eq:wij}
\end{equation}
This should work perfectly well if we can calibrate the $r_{ij}$ matrices properly. For a large
fiducial survey with about 5000 sq. deg. we need about $\simeq 1\%$ accuracy in $r_{ij}$
\citep{Gaztanaga2012}. In practice, the accuracy of the above reconstruction
 can be used to put requirements on the photo-$z$ calibration.

\section{Optimization of the PAU filter set}
\label{sec:opti}

In this section we want to explore how the photo-$z$ performance changes under variations of the PAU NB filter set. 

\subsection{NB filter set variations}
\begin{figure*}
\centering
\includegraphics[height=130mm]{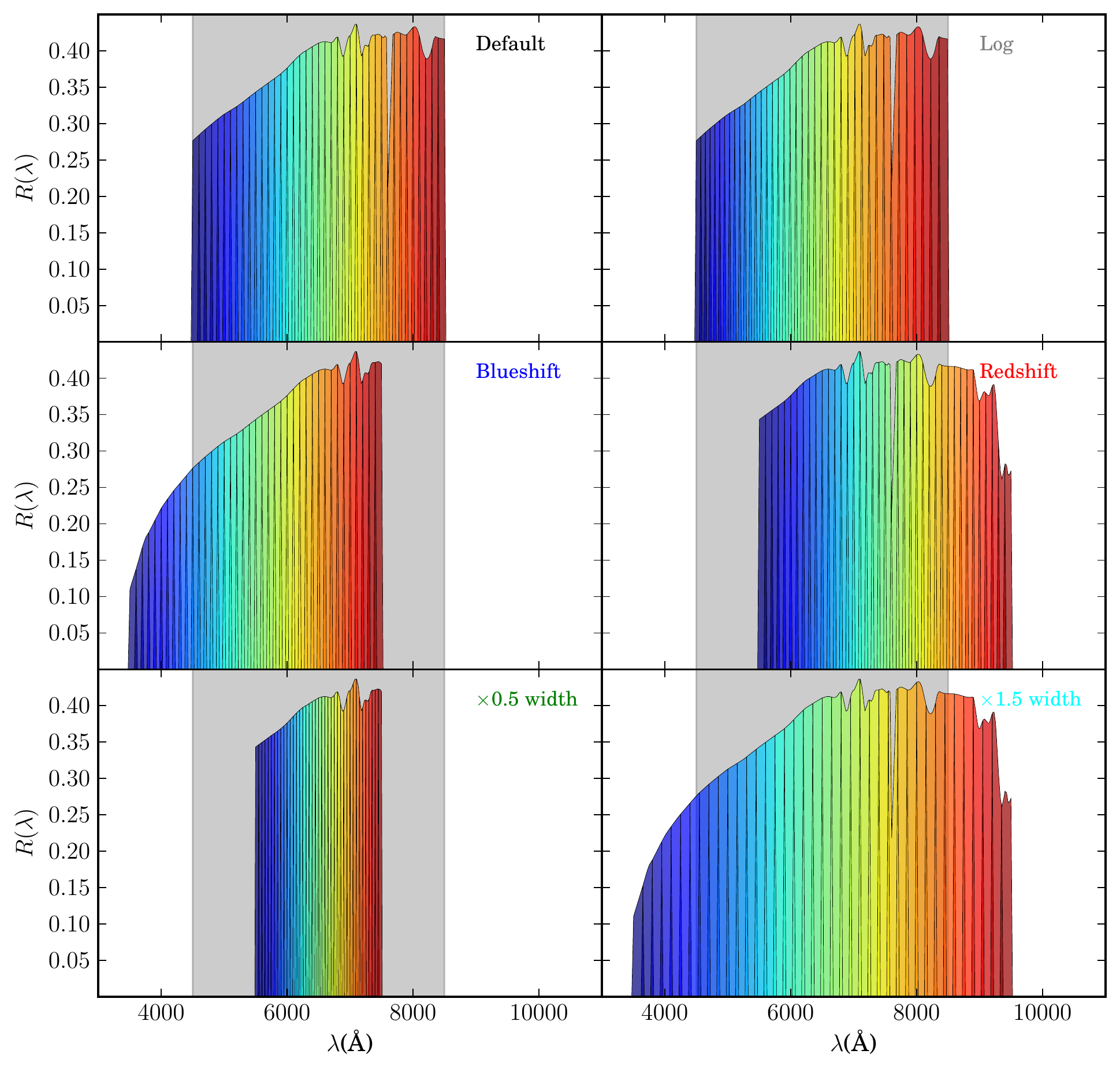}
\caption{On the top-left, the original PAU NB filter set (same as in Fig.~\ref{pau_effective_bands}). The rest are the five variations to be compared in terms of photo-$z$ performance. Grayed areas show the covered wavelength range by the \texttt{Default} filter set. In descending order from left to right we have: the \texttt{Log} filter set with the same overall range as the \texttt{Default} but with band widths that increase logarithmically; the \texttt{Blueshift} filter set, which is the same as the \texttt{Default} but with the bands shifted 1000\AA \ towards bluer wavelengths; the \texttt{Redshift} filter set which is the same as \texttt{Default} but shifted towards redder wavelengths; the \texttt{$\times$0.5 width} filter set whose band widths are half those of the \texttt{Default} ones; and the \texttt{$\times$1.5 width} filter set whose bands are 1.5 times wider. The overall wavelength ranges of these last two set-ups are chosen to be centered respect to the range of the \texttt{Default} set.}
\label{pau_filt_sets}
\end{figure*}

We study five variations of the original NB filter set whose response is shown in Fig.~\ref{pau_filt_sets}. All the variations conserve the number of filters. In the order that appear in Fig.~\ref{pau_filt_sets}, the proposed filter sets are:

\begin{table}
\centering
\caption{Global photo-$z$ performance results for each filter set shown in Fig.~\ref{pau_filt_sets}.  Photo-$z$ performance is characterized through the three metrics: bias (median), $\sigma_z$ ($\sigma_{68}$) and the $3\sigma$-outlier fraction. Photo-$z$ quality cuts resulting in a 50\% \ overall completeness are applied in all cases. We show results for the
Bright Sample (BS) and Faint Sample (FS).}
\begin{tabular}{l|c|c|c|}
 & Bias & $\sigma_z$(\%) & Outliers(\%) \\
\hline
Default BS & -0.71$\cdot10^{-4}$ & 0.34 & 3.02
\\
Default FS & -1.33$\cdot10^{-3}$ & 4.73 & 7.36
\\
\hline
Blueshift BS &-2.11$\cdot10^{-4}$ & 0.38 &  3.23
\\
Blueshift FS & -3.11$\cdot10^{-3}$ & 5.19 & 7.05
\\
\hline
Redshift BS &-0.74$\cdot10^{-4}$  & 0.35 &   3.31
\\
Redshift FS &  -0.65$\cdot10^{-3}$  & 4.99 & 7.21
\\
\hline
Log BS &-0.69$\cdot10^{-4}$  & 0.35 &   2.80
\\
Log FS &  -1.46$\cdot10^{-3}$  & 4.73 & 7.43
\\
\hline
x1.5 width BS &-3.18$\cdot10^{-4}$  & 0.45 &   3.00
\\
x1.5 width FS &  -2.76$\cdot10^{-3}$  & 3.87 & 7.76
\\
\hline
x0.5 width BS &-0.00$\cdot10^{-4}$  & 0.32 &   5.22
\\
x0.5 width FS &  -0.99$\cdot10^{-3}$  & 6.91 & 5.52
\\
\hline
\end{tabular}
\label{tab:pz_results_filt_sets}
\end{table}

\begin{itemize}
\item \textbf{Default:} This is the default filter set already shown in Fig.~\ref{pau_effective_bands}. 
\item \textbf{Log:} In this filter set, band widths increase in wavelength logarithmically, so that they fulfill $\lambda_0 / \Delta \lambda=const.$, where $\Delta \lambda$ is the width of the rectangular part of the band (without taking into account the lateral wings) and $\lambda_0$ is the central wavelength of the band. We impose the overall wavelength range covered by the set of bands to be the same as for the \texttt{Default} filter set. Given that the total number of bands is kept at 40, we obtain that the bluest filter has a width of 97\AA, while the reddest is 159\AA \ wide. The reason for this filter set is that, when spectra are redshifted, their spectral features are moved to redder wavelengths, but also their widths are stretched as $\Delta \lambda ' = (1+z) \Delta \lambda$. If the photo-$z$ determination depends strongly on the tracking of any spectral feature, such as the 4000\AA \ break in elliptical galaxies, a filter set like \texttt{Log} will continue to enclose the same part of the spectral feature in a single feature independently of how redshifted is the spectrum.
\item \textbf{Blueshift:} This is the same as the \texttt{Default} filter set, however bands have been shifted 1000\AA \ towards bluer wavelengths. We expect to get better photo-$z$ performance at low redshift and for late-type galaxies. The down side of this filter set is that the overall response turns out to be very inefficient in the ultraviolet zone (middle-left of Fig.~\ref{pau_filt_sets}), like it was for the \textit{u} band.
\item \textbf{Redshift:} This is the same variation as before but shifting bands 1000\AA \ towards redder wavelengths. We expect to get better photo-$z$ performance at high redshift and for early-type galaxies. This filter set does not suffer from the problem of the ultraviolet, so its band responses are much more uniform over the covered range (middle-right of Fig.~\ref{pau_filt_sets}). On the other hand, the sky brightness on this region is higher.
\item \textbf{$\times$0.5 width:} This is a filter set whose band widths are half of the \texttt{Default} ones. Lateral wings are also reduced to half of their size, from 25\AA \ to 12.5\AA, in order to avoid an excessive overlap between adjacent bands. We expect to improve the photo-$z$ precision, at least for galaxies with good Signal-to-Noise ratio on their photometry. The down side of this filter set is that, since the number of bands is kept, the overall wavelength range covered is also reduced by a half. We choose it to be centered with respect to the \texttt{Default}, so that it covers from 5500\AA \ to 7500\AA, roughly spanning only from the bluest filter of the \texttt{Redshift} set to the reddest of the \texttt{Blueshift} set. This can lead to a degradation of the photo-$z$s at very low and high redshift, although the broad bands may attenuate this effect.
\item \textbf{$\times$1.5 width:} This is a filter set whose band widths are 1.5 times wider than the \texttt{Default} ones. Because of this, we expect a significant degradation of the photo-$z$ precision for galaxies with good Signal-to-Noise ratio on their photometry. However, the increase in $S/N$ may help. Moreover, the covered wavelength range also increases by 50\%. We choose the new range to be centered with respect to the \texttt{Default} set, so that it covers from 3500\AA \ to 9500\AA, roughly from the bluest edge of the \texttt{Blueshift} set to the reddest edge of the \texttt{Redshift} set, so we expect to see a more uniform photo-$z$ performance over the whole redshift range.
\end{itemize}

We generate magnitudes for each filter set band as described in Section~\ref{sec:mock} using the same exposure times per NB filter tray and BB as in the \texttt{Default} filter set. We do not try to optimize the exposure times for each filter set. The aim of this study is to see how, in spite of this, the photo-$z$ performance changes. Once the new photometric mock catalogs are created, they are also split into a Bright Sample ($i_{AB}<22.5$) and Faint Sample ($22.5<i_{AB}<23.7$). We run BPZ on each catalog using the same settings as for the \texttt{Default} filter set. There is no need to calibrate a different prior for each filter set, since the prior was initially calibrated on the broad band $i$, which is shared by all these filter sets. Photo-$z$ quality cuts resulting in an overall completeness of $\sim$50\% are applied in all cases.
\begin{figure*}
\centering
\includegraphics[type=pdf,ext=.pdf,read=.pdf, width=130mm]{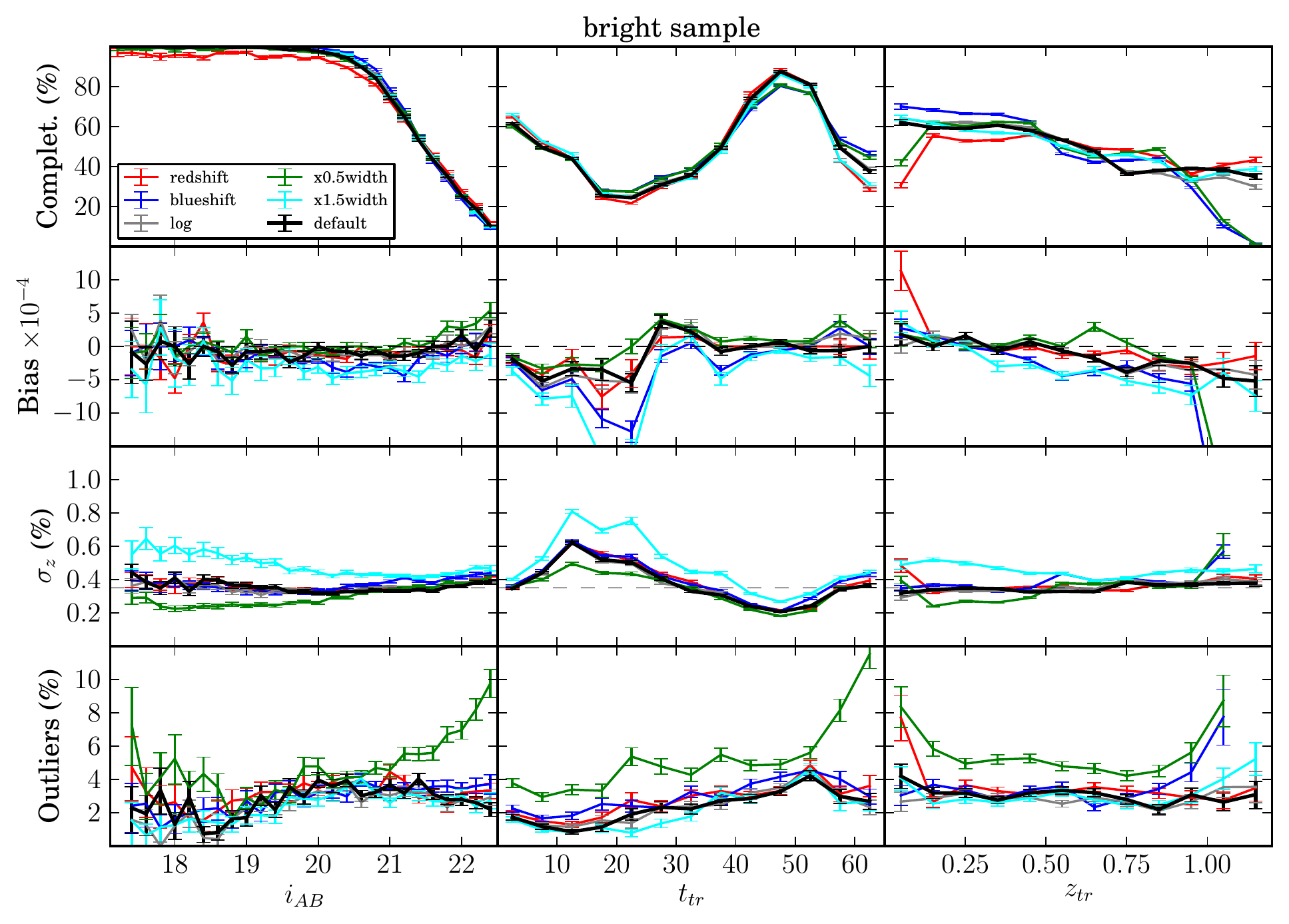} \\
\includegraphics[type=pdf,ext=.pdf,read=.pdf, width=130mm]{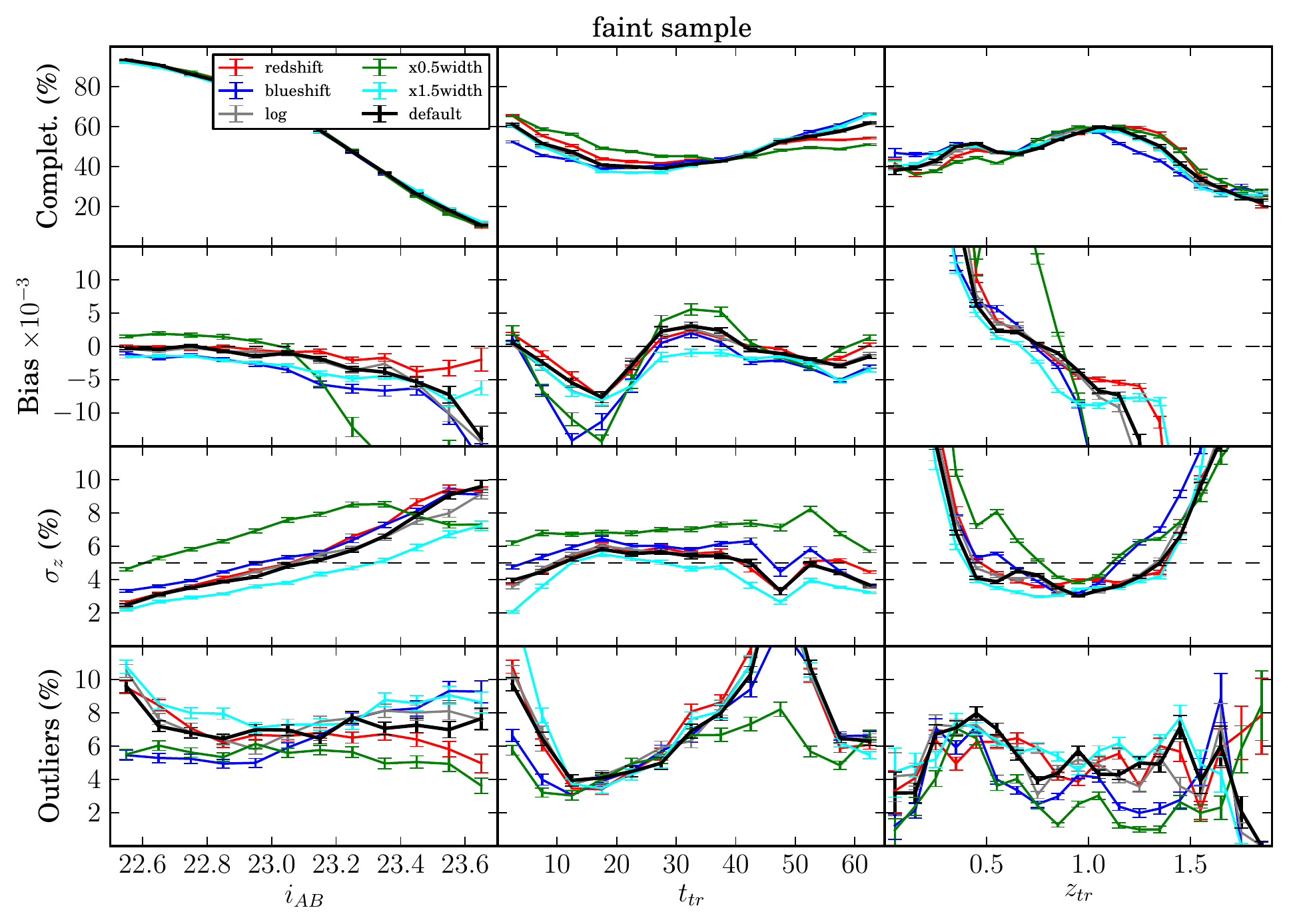}
\caption{Photo-$z$ performance metrics using the different filter sets shown in Fig.~\ref{pau_filt_sets}. Rows: Completeness after applying a photo-$z$ quality cut leading to a 50\% global completeness; bias (median); $\sigma_z$ ($\sigma_{68}$) and 3$\sigma$-outlier fraction, as a function of $i_{AB}$, $t_{tr}$ and $z_{tr}$ (columns), in the BS (top) and the FS (bottom).}
\label{pz_results_filt_sets}
\end{figure*}

\subsection{Global photo-$z$ performance}
Global photo-$z$ performance results for each filter set are shown on Table~\ref{tab:pz_results_filt_sets}, using the same metrics as in Section~\ref{sec:photoz}: bias (median), $\sigma_z$ ($\sigma_{68}$) and the $3\sigma$-outlier fraction. We find that the \texttt{$\times$0.5 width} set gives the best bias (it completely vanishes), and $\sigma_z$ ($\sim$6\% better than \texttt{Default}) in the BS, while in the FS it is the \texttt{Redshift} set which gives the best bias ($\sim$54\% better) and the \texttt{$\times$1.5 width} set which gives the best $\sigma_z$ ($\sim$18\% better). On the other hand, the \texttt{$\times$1.5 width} set gives the worst bias (a factor 4.6 worse) and $\sigma_z$ ($\sim$32\% worse) in the BS, while in the FS it is the \texttt{Blueshift} set which gives the worst bias (a factor 2.4 worse) and the \texttt{$\times$0.5 width} which gives the worst $\sigma_z$ ($\sim$46\% worse). Regarding the outlier fraction, its direct comparison is trickier since it depends on the value of $\sigma_z$. Even so, we see that in the BS the \texttt{Log} set gives the best value, while in the FS the \texttt{$\times$0.5 width} set gives the worst. 

The general conclusions are that the \texttt{Log} set gives almost the same photo-$z$ performance as the \texttt{Default} set, with a slight increase of $3\%$ in $\sigma_z$ in the BS. Therefore, we see that the logarithmic broadening of the band widths does not provide any global improvement. On the other hand, and as we expected, if the Signal-to-Noise ratio in the photometry is good enough, the narrower the bands, the better the photo-$z$ performance results. On the other hand, wider bands are the ones that give better photo-$z$ precision in the FS, because there are more bands that pass the cut $\sigma_m<0.5$ introduced in Section~\ref{sec:mock}. 

\subsection{Results as a function of $i_{AB}$, $t_{tr}$ and $z_{tr}$}
In Fig.~\ref{pz_results_filt_sets} we show plots similar to those in Figs.~\ref{bs_pz_results} and \ref{fs_pz_results} with the photo-$z$ performance metrics as a function of $i_{AB}$, $t_{tr}$ and $z_{tr}$ for the BS (top) and the FS (bottom) when using the different filter sets of Fig.~\ref{pau_filt_sets}. Black curves correspond to the photo-$z$ results of the \texttt{Default} filter set when the 50\% completeness photo-$z$ quality cut is applied, and we will treat them as the reference results. The rest of curves in different colors correspond to the variations of the \texttt{Default} filter set. As a general trend, we see that these curves do not deviate much from the reference. Even so, we will discuss each case separately. 

In the BS, we see that redshifting the bands (red curves) slightly degrades the completeness at low magnitudes up to $i_{AB}<21.5$. As was expected, all the metrics also degrade at low $z_{tr}$. In contrast, blueshifting the bands (blue curves) shows the opposite behavior, a degradation of all the metrics at high $z_{tr}$. This is due to the lack of coverage at blue and red wavelengths respectively of each filter set, as we have already mentioned before. Something similar happens for the \texttt{$\times$0.5 width} set, where band widths, and consequently the covered wavelength range, is reduced by a half (green curve). The resulting photo-$z$ performance is worse at both low and high $z_{tr}$. However, the photo-$z$ precision $\sigma_z$ is slightly better at intermediate redshifts ($0.15<z_{tr}<0.5$), for spiral galaxies ($10<t_{tr}<30$) and at bright magnitudes ($i_{AB}<20.5$). Increasing the band width by a factor 1.5 (cyan curve) does not result in an improvement in any case. Bias and $\sigma_z$ degrade all over the range of the three variables, $i_{AB}$, $t_{tr}$ and $z_{tr}$. This is in full agreement with the results shown in Table~\ref{tab:pz_results_filt_sets}, where this filter set was seen as giving the worst photo-$z$ performance. Also in agreement with Table~\ref{tab:pz_results_filt_sets}, we see that the \texttt{Log} filter set practically does not introduce any change from the \texttt{Default} filter set. 

In the FS we do not observe big differences between the completeness curves, but for example the \texttt{Blueshift} filter set shows slightly lower completeness for elliptical galaxies than for irregulars, unlike the \texttt{Redshift} and \texttt{$\times$0.5 width} filter sets, which show the opposite behavior. In the $z_{tr}$ range we also recognize similar behaviors as in the BS, as for example the fact that the \texttt{Blueshift} filter set shows better completeness at low $z_{tr}$ and worse at high, as well as the opposite behavior of the \texttt{Redshift} and \texttt{$\times$0.5 width} filter sets. The \texttt{Blueshift} filter set seems to cause a significant degradation in the bias for faint, spiral and high redshift galaxies, and also delivers a considerably worse $\sigma_z$ than the \texttt{Default} over all the ranges. In return, the \texttt{Redshift} filter set shows better bias at high magnitudes and redshifts. On the other hand, we observe that the \texttt{$\times$0.5 width} filter set shows much more pronounced trends on the bias, in particular at $i_{AB}>23.1$, spiral galaxies and over all the $z_{tr}$ range, where values are substantially worse than for the \texttt{Default} filter set. As in Table~\ref{tab:pz_results_filt_sets}, we observe that the worst $\sigma_z$ is found for the \texttt{$\times$0.5 width} filter set, while the best is for the \texttt{$\times$1.5 width} filter set over all the ranges. This is exactly the opposite to the behavior seen for the BS. 
In general, narrower bands are useful in the BS, but not in the FS.  

\section{Discussion and Conclusions}
\label{sec:discussion}
In the previous sections, we have seen that, at the level of simulated data, a photo-$z$ precision of $\sigma_z \sim 0.0035(1+z)$ can be achieved for $\sim$50\% of galaxies at $i_{AB}<22.5$ by using a photometric filter system of 40 narrow bands of 125\AA \ width together with the \textit{ugrizY} broad bands. The precision degrades to $\sigma_z \sim 0.05(1+z)$ when we move to the magnitude range $22.5<i_{AB}<23.7$. These coincide with the two photo-$z$ precision requirements defined in \citet{Gaztanaga2012} needed to simultaneously measure Redshift Space Distortions (RSD) and Magnifications bias (MAG) on two samples, one on the foreground and one on the background, over the same area of the sky. The galaxies removed are the ones with the worst photo-$z$ quality according to our photo-$z$ algorithm used. In \citet{Marti2014a} it is shown that this kind of cuts, when they remove a substantial fraction of galaxies, can grossly bias the measured galaxy clustering. However, in the same paper the authors propose a way to correct for it. 

In this analysis we use a set of templates to generate the SEDs for the test galaxies, and then a subset of this same template set to mesure the photo-$z$. This is clearly an idealized process that cannot be used with real data. However, when dealing with real data, one can still modify and optimize the templates used in the photo-$z$ determination so that they reproduce as closely as possible the observed SEDs. Therefore, we consider the results here as reasonably realistic. 

On the other hand, we found that spiral galaxies are the ones that give the worst photo-$z$ performance. Moreover, quality cuts mostly remove them. Contrary to what was assumed in \cite{Benitez2009}, elliptical galaxies do not provide the best photo-$z$ performance, but irregular galaxies with prominent emission lines at $\sim$3737\AA \ [OII] and $\sim$5000\AA \ [OIII] are actually the ones that give the best performance. A possibility is that these two emission lines are better traced by the narrow bands than a single feature as the 4000\AA \ break of elliptical galaxies, making the photo-$z$ determination more robust. 

Furthermore, these lines are narrower than the relatively broad 4000\AA \ break, and this would explain the higher precision observed for irregular galaxies. However, a caveat may be in order here: the variability in the equivalent widths of these lines that is found in nature might not be completely captured by the templates we have used, so that in real data the photo-$z$ precision for these irregular galaxies might degrade slightly.

We also studied the effect of including the 40 narrow bands in a typical broad band filter set \textit{ugrizY}. We find that the $\Delta z / (1+z_{tr})$ distributions become more peaky around the maximum moving away from Gaussianity. Bias improves by an order of magnitude. Precision also improves in a factor of $\sim$5 below $i_{AB}\sim22.5$. However, the low signal-to-noise in narrow bands makes the improvement very small within $22.5<i_{AB}<23.7$, concluding that narrow bands at faint magnitudes are useful to improve the bias but not the precision.

We have also estimated the photo-$z$ migration matrices $r_{ij}$ which correspond to the probability
 that a galaxy observed at the photo-$z$ bin $i$ is actually at the true redshift bin $j$.
These are shown  in Fig.~\ref{plot:rij} for both the BS and FS.  We then show (in Fig.~\ref{plot:wij})
how this photo-$z$ migration matrix $r$ distorts the observed auto and cross-correlation of galaxies in narrow redshift bins.
 We show results for both the intrinsic clustering, which dominates the diagonal in the measured angular 
cross-correlation matrix,  $\bar{\omega}$, and the magnification effect, which appears as a diffused
off-diagonal cloud in $\bar{\omega}_{ij}$. The true cross-correlation matrix $\omega_{ij}$ 
can be obtained from the inverse migration $r^{-1}$  with a simple matrix operation: 
$\omega = r^{-1}\cdot\bar{\omega}\cdot(r^T)^{-1}$. This is a standard deconvolative problem, and it is only limited
by how well we know the migration matrix.

In the last section, we find that introducing slight variations on the 40 narrow band filter set, such as: shifting bands to the higher/lower wavelengths, narrowing/broadening or increasing logarithmically band widths, do not introduce significant changes on the final photo-$z$ performance.  Even so, general trends are that narrowing/broadening bands improve/worses the photo-$z$ performance below/above $i_{AB}\sim22.5$, while red/blueshifting bands improve the photo-$z$ performance at high/low redshifts or the quality-cuts efficiency for early/late-type galaxies. Logarithmically growing band widths do not turn into any measurable improvement. Therefore, we conclude that the initial proposed filter set of 40 narrow bands seems to be close to optimal for the purposes of the PAU Survey at the WHT.


\section*{Acknowledgments} 
We thank Christopher Bonnett, Ricard Casas, Samuel Farrens, St\'{e}phanie Jouvel, Eusebio S\'{a}nchez and Ignacio Sevilla for their help and useful discussions.
Funding for this project was partially provided by the Spanish
Ministerio de Econom\'{\i}a y Competitividad (MINECO) under projects
AYA2009-13936, AYA2012-39559, AYA2012-39620, FPA2012-39684,
Consolider-Ingenio 2010 CSD2007-00060, and
Centro de Excelencia Severo Ochoa SEV-2012-0234.

\bibliography{../bibtex/library}

\label{lastpage}

\bibliographystyle{mn2e}

\end{document}